\documentclass[11pt,preprint]{aastex}

\newcommand{\HII}{H {\small II}  }
\newcommand{\kms}{{\rm ~km~s}^{-1}}
\newcommand{\Msun} {M_\sun}

\newcommand{\mjyb}{{\rm ~mJy~beam}^{-1}}

\newcommand{\column}{cm$^{-2}~$}

\newcommand{\mjybkms}{{\rm mJy~beam}^{-1}{\rm ~km~s}^{-1}}
\newcommand{\jybkms}{{\rm Jy~beam}^{-1}{\rm ~km~s}^{-1}}

\shorttitle{FROM CONVERGENCE OF FILAMENTS TO DISK-OUTFLOW ACCRETION}
\shortauthors{GALV\'AN-MADRID ET AL.}

\begin{document}


\title{FROM THE CONVERGENCE OF FILAMENTS TO DISK-OUTFLOW ACCRETION: MASSIVE-STAR FORMATION IN W33A}


\author{Roberto Galv\'an-Madrid\altaffilmark{1,2,3},  
Qizhou Zhang\altaffilmark{1}, Eric Keto\altaffilmark{1}, \\
Paul T. P. Ho\altaffilmark{1,3}, Luis A. Zapata\altaffilmark{2,4}, 
Luis F. Rodr\'iguez\altaffilmark{2},\\ Jaime E. Pineda\altaffilmark{1}, 
and Enrique V\'azquez-Semadeni\altaffilmark{2}}

\email{rgalvan@cfa.harvard.edu}


\altaffiltext{1}{Harvard-Smithsonian Center for Astrophysics, 60 Garden Street, 
Cambridge MA 02138, USA}
\altaffiltext{2}{Centro de Radioastronom{\'\i}a y Astrof{\'\i}sica, Universidad 
Nacional Aut\'onoma de Mexico, Morelia 58090, Mexico}
\altaffiltext{3}{Academia Sinica Institute of Astronomy and Astrophysics, P.O. 
Box 23-141, Taipei 106, Taiwan}
\altaffiltext{4}{Max-Planck-Institut f\"{u}r Radioastronomie, Auf dem H\"{u}gel 69,53121, Bonn, Germany}


\slugcomment{ApJ in press}

\begin{abstract}

Interferometric observations of the W33A massive star-formation region, performed 
with the Submillimeter Array (SMA) and the Very Large Array (VLA) at resolutions from 
5\arcsec ~ (0.1 pc) to 0.5\arcsec ~  (0.01 pc) are presented.  Our three main findings are: 
(1) parsec-scale, filamentary structures of cold molecular gas are detected. 
Two filaments at different velocities intersect in the zone where the star formation 
is occurring. This is consistent with 
triggering of the star-formation activity by the convergence of such filaments, as  
predicted by numerical simulations of star formation initiated by converging flows. (2) The two 
dusty cores (MM1 and MM2) at the intersection of the filaments  are found to be at different 
evolutionary stages,  
and each of them is resolved into multiple condensations. MM1 and MM2 have markedly different 
temperatures, continuum spectral indices, molecular-line spectra, and masses of both stars and gas. 
(3) The dynamics of the ``hot-core'' MM1 indicates the presence of a rotating disk in its center 
(MM1-Main) around a faint free--free source. The stellar mass is estimated 
to be $\sim 10~\Msun$. A massive molecular outflow is observed along the rotation axis of the 
disk.

\end{abstract}


\section{Introduction} \label{intro}

Stars form by accretion of gas in dense molecular-cloud cores.  
However, the differences, if any, in the details of the formation process of massive stars
(those with roughly $M_\star > 8 ~ M_\odot$) compared to low-mass stars are not well understood.  
Recent reviews on the topic are those by \cite{Beu07}, and \cite{ZY07}. 

We are carrying out a program aimed at studying how the formation of massive stars in clusters 
proceeds in the presence of different levels of ionization, from the onset of detectable 
free--free emission to the presence of several bright ultracompact (UC) \HII regions. 
In this paper we present our first 
results on the massive star-formation region W33A (also known as G12.91-0.26), at a kinematic distance of 
3.8 kpc \citep{Jaffe82}. 
W33A is part of the W33 giant \HII region complex \citep{W58}. It was recognized 
as a region with very high far infrared luminosity ($\approx 1 \times 10^5$ $L_\odot$ ), 
but very faint radio-continuum emission by \cite{Stier84}. 
\cite{vdT00} 
modeled the large-scale (arcminute) cloud as a spherical envelope with a 
power-law density gradient, based on single-dish mm/submm observations. 
Those authors also presented mm interferometric observations at several-arcsecond resolution that 
resolved the central region into two dusty cores separated by $\sim 20,000$ AU. The brightest mm core 
contains faint ($\sim 1$ mJy at cm wavelengths) radio-continuum emission \citep{RH96}
resolved at 7 mm into possibly three sources separated by less than 
$1\arcsec$ ($\approx 4000$ AU) from each other \citep{vdTM05}. 
These radio sources were interpreted by \cite{vdTM05} as the gravitationally trapped \HII regions 
set forth by \cite{Keto03}.  
However, the earlier detection by \cite{Bunn95} of near-infrared recombination 
line (Br$\alpha$)  emission with FWHM $=155~\kms$ suggests that at least some of the 
radio free--free emission is produced by a fast ionized outflow. More recently, 
\cite{Davies10} reported spectroastrometry observations of Br$\gamma$ emission toward W33A. 
The Br$\gamma$ emission appears to be produced by at least two physical components: broad line wings 
extending to a few hundreds of kilometers per second from the systemic velocity appear to trace a bipolar jet 
on scales of a few AU, while the narrow-line emission may be attributed to a dense 
\HII region \citep{Davies10}.  
Being a bright mid- and far-infrared source, W33A has also been target of 
interferometry experiments at these wavelengths, which reveal density gradients and non-spherical 
geometry in the warm dust within the inner few hundred AU \citep{dW07,dW09}. 

Here we report on millimeter and centimeter interferometric observations performed with the 
Submillimeter Array (SMA) and the Very Large Array (VLA) at angular resolutions 
from $\sim 5\arcsec$ to $0.5\arcsec$. We find a massive star-forming cluster embedded  
in a parsec-scale filamentary structure of cold molecular gas. The dense gas is 
hierarchically fragmented into two main dusty cores, each of them resolved into more peaks 
at our highest angular resolution. The main cores appear to be at different evolutionary 
stages, as evidenced from their differing spectra, masses, temperatures, and continuum spectral 
indices. 
The warmer core harbors faint free--free emission centered on a rotating disk traced by warm molecular gas. 
The disk powers a massive molecular outflow, indicating active accretion. 
In Section 2 of this paper, we 
describe the observational setup. In Section 3 we list our results, in Section 4 we present a discussion of 
our findings, and in Section 5 we give our conclusions.

\section{Observations} \label{obs}

\subsection{SMA} \label{sma}

We observed the W33A region with the
Submillimeter Array\footnote{The Submillimeter Array is a joint project between the Smithsonian 
Astrophysical Observatory and the Academia Sinica Institute of Astronomy and Astrophysics 
and is funded by the Smithsonian Institution and the Academia Sinica.} 
\citep{Ho04}  
in the 1.3-mm (230 GHz) band using two different array configurations. Compact-array 
observations were taken on 2007 July 17, and covered baselines with lengths between 7 
and 100 K$\lambda$ (detecting spatial structures in the range of $29.5\arcsec$ to $2.1\arcsec$). 
Very Extended (VEX) configuration data were taken on 2008 August 2, with baseline lengths 
from 23 to 391 K$\lambda$ ($9.0\arcsec$ to $0.5\arcsec$). For both observations, the  
two sidebands covered the frequency ranges of $219.3-221.3$ and $229.3-231.3$ GHz 
with a uniform spectral resolution of $\approx 0.5$ $\kms$.  

We also report on the continuum emission from archival observations taken in the 0.9-mm (336 GHz) 
band on 
2006 May 22. The array was in its Extended configuration, with baseline lengths from 18 to 232 
K$\lambda$ ($11.4\arcsec$ to $0.8\arcsec$). 
These data were used to constrain the spectral index of the continuum sources.

The visibilities of each data set were separately calibrated using the
SMA's data calibration program, MIR.   
We used Callisto to obtain the absolute amplitude and quasars to derive the  
time-dependent phase corrections and frequency-dependent bandpass corrections. 
Table 1 lists relevant information on the calibrators. 
We estimate our flux-scale uncertainty to be better than $15 \%$. 
Further imaging and processing was done in MIRIAD and AIPS. 

The continuum was constructed in the $(u,v)$ domain from the line-free channels. The line-free 
continuum in the 1.3-mm Compact-configuration data was bright enough to perform phase self-calibration. 
The derived gain corrections were applied to the respective line data. No self-calibration 
was done for the higher angular resolution data sets.

\subsection{VLA} \label{vla}

We observed the $(J,K)=(1,1)$ and $(2,2)$ inversion transitions of NH$_3$ 
with the 
Very Large Array\footnote{
The National Radio Astronomy Observatory is operated by Associated Universities, Inc., 
under cooperative agreement with the National Science Foundation.}. 
Observations were carried out on 2004 June 14 and 15 (project AC733). The array 
was in its D configuration, with baseline lengths in the range of 3 to 79 $K\lambda$ 
(detecting scales from 68.7\arcsec to 2.6\arcsec). The correlator was set to the 4-IF mode. 
Each of the IF pairs was tuned to the (1,1) and (2,2) lines, respectively, 
covering a bandwidth of 3.1 MHz (39 $\kms$) at a spectral resolution of 0.6 $\kms$.

The data were calibrated and imaged using standard procedures in the AIPS software. 
Table 1 lists the quasars used to derive the absolute flux scale, the time-dependent 
gain corrections, and the frequency-dependent passband calibration. The absolute 
flux scale is accurate within a few percent.  
No self-calibration was performed.

\section{Results} \label{res}

\subsection{Continuum Emission} \label{cont}

\subsubsection{Morphology} \label{mor}

\begin{figure}
\figurenum{1}
\begin{center}
\plottwo{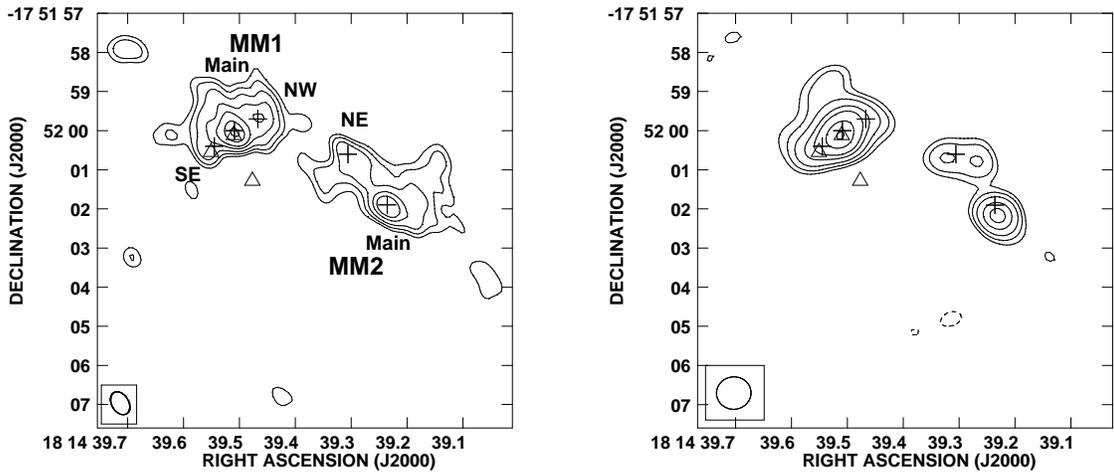}{f1b.eps}
\caption{
(Sub)millimeter continuum emission in W33A. 
The {\it left} panel shows the 231 GHz 
(1.3 mm) continuum from the SMA Compact+VEX data (HPBW $= 0.63\arcsec \times 0.43 \arcsec, 
P.A. = 30.7^\circ$). 
Contours are at $-5,5,7,10,15,20,30,$ and 40 times 
the noise of 1.5 $\mjyb$. The {\it right} panel shows the 336 GHz (0.9 mm) continuum from 
the Extended-configuration data (HPBW $= 0.88\arcsec \times 0.83 \arcsec, P.A. = 275.1^\circ$), 
with contours at $-5,5,7,10,15,20,30,$ and 39 times the rms noise of 6 $\mjyb$. 
The cores MM1 and MM2 are labeled, and the sources into which they fragment are 
marked by {\it crosses}. {\it 
Triangles} mark the positions of the faint 7 mm sources reported by van der Tak \& Menten 
(2005). 1 arcsec corresponds to 3800 AU (0.018 pc).
}
\end{center}
\label{fig1}
\end{figure}

Our observations at 1.3 mm resolve each of the two 
mm cores reported by \cite{vdT00} into multiple continuum sources. The concatenation 
of Compact- and VEX-configuration data permits us to simultaneously resolve 
the structures at $\approx 0.5\arcsec$ resolution and to be sensitive to relatively extended structures.  
Figure \ref{fig1} ({\it left}) shows the 1.3-mm continuum map. 
It is seen that MM1 and MM2 are resolved into 
at least three and two smaller mm peaks respectively (marked by crosses in Fig. \ref{fig1}). 
Only MM1 is associated with the cm emission detected by \cite{RH96}. 
Two of the three faint 7-mm sources ($S_\mathrm{7mm}\sim 1$ mJy) 
reported by \cite{vdTM05} toward MM1 at a resolution of $\sim 0.05\arcsec$ 
(marked by triangles in Fig. \ref{fig1}) are counterparts of the 1.3-mm peaks. 
The faintest 7-mm source has no association in our continuum or line data. 
In addition to the clearly identified 1.3-mm peaks, the northeast-southwest large-scale 
continuum ridge appears to have more fainter sources. Another possible source is well separated 
from the ridge, at $\approx 8\arcsec$ to the southwest of MM1. More sensitive observations 
are necessary to investigate their nature. 

We label the identified mm peaks as MM1-Main (the brightest source of MM1), MM1-NW (for northwest), 
MM1-SE (for southeast), MM2-Main, and MM2-NE. 
Table 2 lists the peak positions and peak intensities measured in the mm map of 
Fig. \ref{fig1} {\it left}.  
The sums of the 1.3-mm fluxes of the components that we obtain from multi-component Gaussian fits 
to the sources comprising MM1 and MM2 are robust, and consistent with the fluxes measured by 
integrating the intensity over the areas of interest.  However, the sizes and fluxes   
of the individual components in the fits are not accurate, mainly because of insufficient angular 
resolution.  
Table 2 lists the added flux of the subcomponents of MM1 and MM2. The ratio of the 
1.3-mm flux of MM1 to that of MM2 in our data is 1.2, very close to that 
reported by \cite{vdT00}: 1.3. The fluxes that we report are $88 \%$ to $106 \%$ larger than 
those in \cite{vdT00}, probably due to differences in $(u,v)$ coverage and flux-scale 
uncertainties.

Only the bright, compact sources are detected in the 0.9-mm continuum image 
(Fig. \ref{fig1} {\it right}). 
This single-configuration data set has a more modest $(u,v)$ coverage than the concatenated 
1.3-mm data and is less sensitive to extended structures. 

\subsubsection{Nature of the Continuum} \label{sed}

To set an upper limit to the free--free contribution at 1.3-mm  
we extrapolate the 8.4 to 43.3-GHz free--free spectral index $\alpha=1.03\pm0.08$  
(where the flux goes as $S_\nu\propto\nu^\alpha$), calculated from 
the fluxes reported by 
\cite{vdTM05} and \cite{RH96}. 
This is a reasonable assumption since $\alpha$ for free--free sources with moderate 
optical depths, either jets  \citep[e.g.,][]{Brooks07,Hofner07,GH07} or \HII regions 
\citep[e.g.,][]{Franco00,KZK08,GM09}, is approximately in the range from 0.5 to 1.

For MM1 with a 7-mm flux of $S_\mathrm{7mm}\approx4.3$ mJy, the maximum free--free flux at 1.3-mm 
is 28 mJy. 
The 1.3-mm flux integrated over MM1 is $S_\mathrm{1.3mm}\approx357\pm25$ mJy (Table 2). 
Therefore, the free--free contribution to the 1.3-mm flux is at most $\sim 8$ \%. No cm continuum 
has been detected toward MM2 or in the rest of the field, thus, the 1.3-mm 
emission outside MM1 is most probably produced entirely by dust. 
Using the same considerations, the free--free flux of MM1 at 0.9 mm is less than 
$42$ mJy. The integrated flux of MM1 at this wavelength is $S_\mathrm{0.9mm}\sim612$ mJy, 
then the free--free emission is at most $\sim 7$ \%. 
Because the data are taken at different epochs, the possibility of radio variability 
\citep[see][for reports on other targets]{FHR04,vdT05,GM08} adds to the uncertainty. 
The 0.9-mm data may suffer from missing flux, making the fractional 
free--free contribution at this wavelength even smaller. 
In the rest of the paper, we consider the (sub)mm free--free emission to be negligible.

To compare the 0.9-mm and 1.3-mm fluxes in a consistent way,  
we produced images with a uniform $(u,v)$ coverage (using only baselines with lengths from 
30 to 230 $K\lambda$), without self-calibration (using only the VEX data at 1.3 mm and the 
Extended data at 0.9 mm), and a common circular synthesized beam (HPBW $= 0.85\arcsec$). 
The average spectral indices of the two main mm cores are $\langle \alpha_\mathrm{MM1} \rangle = 3.3\pm0.3$ 
and 
$\langle \alpha_\mathrm{MM2} \rangle = 2.5\pm0.4$. In the Rayleigh--Jeans (R-J) approximation ($h\nu \ll k_BT$), 
the spectral index of thermal dust emission is  $\alpha = 2 +\beta$, where 
$\beta$ is the exponent of the dust absorption coefficient. 
The fiducial interstellar--medium (ISM) 
value of $\beta$ is 2, while for hot cores in massive star-forming regions (MSFRs) 
typical values are $\beta\approx1-2$ (e.g., Cesaroni et al. 1999, Zhang et al. 2007). 
Therefore, MM1 has $\beta \approx 1.3$ typical of a hot core, but MM2 has $\beta \approx 0.5$. 
In Section \ref{molec} we show that the 
kinetic temperature of MM2 is $\approx 46$ K, then the R-J limit is not a good 
approximation at 0.9 mm for MM2. 

Without using the assumption of being in the R-J limit, the gas mass $M_\mathrm{gas}$ derived from 
optically--thin dust emission at 1.3-mm can be obtained from Kirchhoff's law: 

\begin{equation}
\Biggl[
\frac{M_\mathrm{gas}}{M_\odot}
\Biggr]
=(26.6) \times \Biggl(\exp\biggl(\frac{11.1}{[T_\mathrm{dust}/\mathrm{K]}}\biggr)-1\Biggr) 
\times 
\Biggl(
\frac{[F_\mathrm{1.3mm}/\mathrm{Jy}][d/\mathrm{kpc}]^2}{[\kappa_\mathrm{1.3mm}/\mathrm{cm}^2\mathrm{g}^{-1}]}
\Biggr), 
\end{equation}

\noindent
where $T_\mathrm{dust}$ is the dust temperature, $F_\mathrm{1.3mm}$ is the 1.3-mm flux density, $d$ is 
the distance to the object, and $\kappa_\mathrm{1.3mm}$ is the dust absorption coefficient. 
Assuming coupling between gas and dust, the dust temperature in MM1 is $\approx 347$ K in the inner 
arcsecond (obtained from fits to CH$_3$CN lines, see Section \ref{innermost}) and $>100$ K at larger scales 
(obtained from NH$_3$ lines, see Section \ref{parsec}). For this range of temperature, 
using an opacity $\kappa_\mathrm{1.3mm}=0.5$ cm$^2$ g$^{-1}$ \citep{OH94}, 
Equation (1) gives a mass for MM1 in the range $M_\mathrm{MM1}=[9,32]~\Msun$.  
For MM2 with a temperature of 46 K (Section \ref{parsec}), the mass is 
$M_\mathrm{MM2} \sim 60~\Msun$.  MM2 then appears to be much colder and more massive (in gas) than 
MM1. The uncertainties in opacity make the mass estimation accurate to only within a factor of a few.

\subsection{Molecular Line Emission} \label{molec}

\subsubsection{The Parsec-scale Gas} \label{parsec}

The large-scale gas within  an area of $\sim 1\arcmin \times 1\arcmin$ 
(or $\sim 1$ pc) can be divided into quiescent gas and  
high-velocity gas. The quiescent gas is best traced by the VLA NH$_3$ data. The high-velocity 
gas is seen in the SMA CO (2--1) maps. 

\subsubsubsection{Morphology and Velocity Structure}

\begin{figure}
\figurenum{2a}
\begin{center}
\includegraphics[angle=0,scale=0.75]{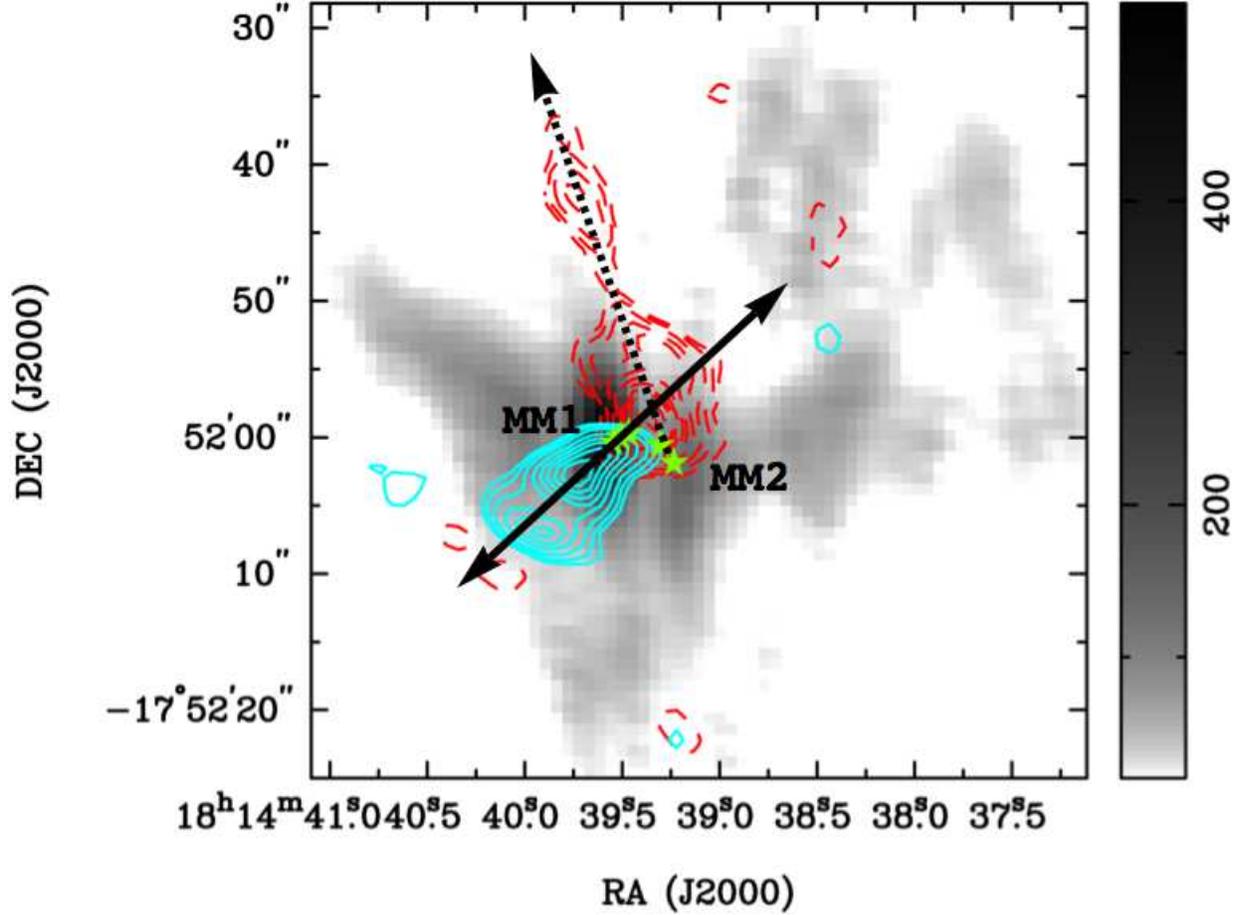}
\end{center}
\caption{
Parsec-scale gas structure toward W33A. The gray scale shows the 
integrated intensity (moment 0, in $\mjybkms$) of the 
quiescent  NH$_3$ (2,2) gas (HPBW $=6.0\arcsec\times2.6\arcsec$, 
P.A. $=2^\circ$). The blue solid contours show the high-velocity gas as detected in CO (2--1) 
(HPBW= $3.0\arcsec\times2.0\arcsec$, P.A. $=56^\circ$) integrated in the range [0,22] $\kms$. 
The red dashed contours show the high-velocity CO gas in the range [62,98] $\kms$. 
Contour levels are $-5,5,7,10,15,20,25,30,35,40,50,$ and $60\times0.7$ $\jybkms$. 
The mm continuum 
sources identified in this paper are marked with stars. 
The directions of the identified outflows are marked with 
arrows. 10 arcsec corresponds to 38,000 AU (0.184 pc). 
}
\label{fig2}
\end{figure}

\indent
º
There is a 
clear morphological difference between the quiescent and the high-velocity gas. 
Figure \ref{fig2} shows moment maps of the NH$_3$ (2,2) line overlaid with the high-velocity CO gas. 
The NH$_3$ moment maps were integrated in the $[31,43]$ $\kms$ LSR velocity range. The blueshifted CO gas 
was integrated in the range $[0,22]$ $\kms$, and the redshifted CO was integrated in $[62,98]$ $\kms$. The 
systemic velocity of the gas closer to MM1 is $V_{sys}\approx 38.5~\kms$ (Section \ref{innermost}). 
The quiescent NH$_3$ emission is composed of one prominent filamentary structure in the east-west direction 
that peaks toward the MM1 region (Fig. \ref{fig2}), 
plus another filamentary structure that extends to the south of MM1 and MM2, 
and some fainter clumps toward the northwest of MM2. The high-velocity CO traces at least two 
molecular outflows that expand outward off the quiescent filaments.  
The lobes of the most prominent 
outflow are centered in MM1, and extend toward the northwest (redshifted gas) and southeast 
(blueshifted) at a position angle P.A. $\approx133^\circ$. The observed size of this 
outflow is about 0.4 pc. The redshifted lobe of a second 
high-velocity outflow extends $\approx 0.5$ pc to the north-northeast 
of the cores at P.A. $\approx19^\circ$, and appears to be originated in MM2. 
The blueshifted side of this outflow does 
not appear at high velocities. The P.A. that we find for the main outflow agrees very well with 
the P.A. $\approx135^\circ$ of the outflow as seen at  2.2 $\mu$m reported by 
\cite[][see their Figure 1]{Davies10}, which matches an elongated $4.5$ $\mu$m structure in $Spitzer$ 
images \citep[Figure 12 of][]{Cyga08}. 
However, the infrared emission is three to four times larger. 
Also, using single-dish observations, \cite{dW09} reported a CO $J=3-2$ outflow whose orientation 
matches those of both the SMA and the near IR outflows. 
The low-velocity CO (2--1) gas could not be properly imaged because 
of the lack of short $(u,v)$ spacings\footnote{Single-dish data were taken for this purpose but 
these were corrupted due to a bad off position and could not be used.}. 

The large-scale quiescent gas has two velocity components, which appear to be two different 
structures of gas at different velocities, as can be seen in Fig. 2b. 
The gas associated with MM1 and MM2 (center of the main filament), as well as the western part of 
the main filament and the north--south extensions appears to be at $\approx 38.5~\kms$, with a typical 
mean-velocity dispersion of $0.4~\kms$.  
The eastern part of the main filament appears blueshifted, with a mean centroid 
velocity of $\approx35.9~\kms$ and a dispersion of the centroid velocity of $0.4~\kms$. 
The two gas structures overlap in space 
toward the MM1/MM2 region, which suggests that the star formation activity in these mm cores 
was triggered by the convergence of the filaments of molecular gas. 
Fig. \ref{fig3} shows a three-dimensional rendering of the same data that better illustrates this result. 
At the position of the mm cores, the filamentary structures do not merely superpose in 
position-position-velocity space, but merge into a region that suddenly extends to 
higher velocities. The larger velocity range at the center (see also Fig. 2c) is found to be due 
to coherent velocity structures (disk and outflows in the dense gas) with the SMA data 
(Section ~ \ref{innermost}).

\subsubsubsection{Physical Parameters} 

\indent

Now we derive the temperature structure of the parsec-scale filaments and lower limits to 
the outflow parameters. 

Most of the gas in the pc-scale filaments, including the gas associated with MM2, is cold, 
with a kinetic temperature $T_\mathrm{kin}=[20,50]$ K.  
$T_\mathrm{kin}$ rises significantly only toward MM1.  
A temperature map at the resolution of the NH$_3$ data (Fig. \ref{fig4}) 
was obtained by fitting the (1,1) and (2,2) line profiles as described in \cite{Roso08}. 
The errors in the temperature determination are in general $\sim 3$ K, 
but get too large toward MM1\footnote{The (2,2) to (1,1) ratio is not sensitive to temperatures much larger 
than 50 K, but we confirm the large temperatures in MM1 at smaller scales using the CH$_3$CN lines, 
see Section ~ \ref{innermost}. The errors in the fits also increase toward MM1 due to its wider 
velocity structure.}. We determine the temperature of MM2 to be  
$T_\mathrm{MM2}\approx46$ K. The temperature of MM1 is constrained to $T_\mathrm{MM1}>100$ K. 

For the CO (2--1) line, the interferometric data suffer from missing flux for the more extended 
emission close to $V_\mathrm{sys}$. We set the following limits to the outflow parameters: 
mass $M_\mathrm{out}>27~\Msun$, 
momentum $P_\mathrm{out}>233~\Msun\kms$, and kinetic energy $E_\mathrm{out}>3\times10^{46}$ erg s$^{-1}$, 
where we corrected for the optical depth at each velocity bin using the $^{13}$CO (2--1) line. 
We  refer the reader to \cite{Qiu09} for a description of the procedure to calculate the aforementioned 
quantities. 
The momentum and especially the energy estimations are less affected by missing 
flux since they depend more on the high-velocity channels.  
\cite{dW09} estimated the inclination 
angle of the inner-cavity walls of the outflow to be $i\sim50^\circ$, by radiative-transfer modeling 
of the mid-IR emission. Correcting by inclination, the lower limits to the momentum and 
energy of the outflow are $P_\mathrm{out}>362~\Msun\kms$ and $E_\mathrm{out}>7\times10^{46}$ erg s$^{-1}$.

\subsubsection{The Inner 0.1 pc} \label{innermost}

\subsubsubsection{Morphology and Velocity Structure}

\indent

The SMA data permit us to study the 
molecular gas at a resolution of $\approx 1500$ AU ($0.4\arcsec$). 
Some molecular lines trace relatively cold gas, while some other lines trace the warmer gas closer 
to the heating sources. Figure \ref{fig5} shows the spectra over the entire sidebands 
from the pixels at the 
positions of the mm peaks MM1-Main and MM2-Main. The prominent lines are labeled, and listed 
along with their upper-level energy in Table 3. Lines with a peak intensity below 20 K are not 
listed. A complete inventory of the molecular lines in W33A will be presented in the future.  
It is immediately seen that MM1 has 
a ``hot-core'' spectrum, while MM2 is almost devoid of molecular emission, if not for the 
CO, $^{13}$CO and C$^{18}$O $J=2-1$, and faint SO $J(K)=6(5)-5(4)$ emission. 
We interpret this difference 
as a signature of the evolutionary stage of the cores, MM1 being more evolved than MM2.

Figure \ref{fig6} shows the moment maps for three lines that exemplify what was mentioned 
above.  The SO 6(5)--5(4) line ({\it top} row) extends in a  
ridge of $\approx 0.1$ pc long in the northeast--southwest direction, from MM1 to MM2. 
The emission is stronger toward MM1 and peaks in MM1-Main. Part of the emission toward MM2 is  
redshifted by $\sim 1-2$ $\kms$ with respect to the emission in the MM1 side, but there is 
no clear velocity pattern. Lines such as SO likely have large optical depths and trace only the 
surface of the emitting region, where clear velocity gradients, especially of rotation, 
may not be expected. From the SO data we constrain any velocity difference between the MM1 and MM2 
cores to $\Delta V<2$ $\kms$. 

For a given molecule, the isotopologue lines and the lines with 
upper energy levels well above 100 K trace the more compact gas, closer to the heating sources. 
Figure \ref{fig6} shows the examples of  $^{13}$CS $J=5-4$ ({\it middle} panel) and CH$_3$CN 
$J(K)=12(3)-11(3)$ ({\it bottom}). 
Both of them are only visible toward MM1, and peak in MM1-Main. These lines trace 
a clear velocity gradient centered on MM1-Main, the blueshifted emission is toward the 
southwest, and the redshifted emission toward  the northeast, perpendicular to the main bipolar CO 
outflow. We interpret this as rotation. 
The emission tracing this velocity gradient is not isolated, there is 
also emission coming from MM1-NW and MM1-SE. Especially in the CH$_3$CN lines, this extra 
emission appears to trace redshifted and blueshifted emission respectively. One possibility is 
that MM1-NW and MM2-SE are separate protostars from MM1-Main. However, the orientation 
of the lobes in the high-velocity outflow is the same. Therefore, we prefer the interpretation that 
MM1-NW and MM2-SE are not of protostellar nature, but emission enhancements (both in continuum 
and line emission) from the hot base of the powerful molecular outflow driven by 
the disk-like structure surrounding MM1-Main. 
In this scenario, the other 7-mm sources reported by \cite{vdTM05} 
(or at least the counterpart of MM1-NW) can be interpreted 
as shocked free--free enhancements in a protostellar jet, similar to those observed in the 
high-mass star formation region IRAS $16547-4247$ \citep{Rod08, FH09}.

Figure \ref{fig7} shows the position--velocity (PV) diagrams of the CH$_3$CN $K=3$ line shown in 
Fig. \ref{fig6},  
centered at the position of MM1-Main perpendicular to the rotation axis ({\it top} frame) and along  
it ({\it bottom} frame).  The rotation pattern is similar to those observed in objects that 
have been claimed to be Keplerian disks, i.e., structures where the mass of the central 
object is large compared to the mass of the gas, rotating with a velocity $V_\mathrm{rot} \propto 
r^{-0.5}$ \citep{Zhang98b,Cesa05,JS09}. 
The large velocity dispersion closest to MM1-Main (Figs. \ref{fig6} and \ref{fig7}) ought to be caused by 
unresolved motions in the inner disk, since velocity dispersions well above 1 $\kms$ cannot be due 
to the gas temperature. 

Recently, \cite{Davies10} reported a possible disk-jet system centered in W33A MM1-Main. 
The jet, observed in the Br$\gamma$ line, extends up to $\pm300~\kms$ in velocity at scales $\sim $ 1 AU, 
with a similar orientation and direction to the molecular outflow reported in this paper. However, the 
velocity structure of what \cite{Davies10} interpret as a rotating disk has a similar orientation but 
opposite sense of rotation as the disk that we report. They used CO absorption lines with upper energy 
levels $E_u\sim 30$ K, while we use emission lines like those of CH$_3$CN, with $E_u>100$ K (Table 3). 
If an extended screen of cold gas with a negligible velocity gradient is between the observer and 
the inner warm gas with a velocity gradient, it is possible that the absorption lines are partially 
filled with emission, mimicking a velocity gradient with the opposite sense than that seen in 
the emission lines.

\subsubsubsection{Physical Parameters}

\indent 

Now we derive the temperature and column density of the hot-core emission, and constrain the 
stellar mass, gas mass, and CH$_3$CN abundance in MM1. 

The kinetic temperature of the innermost gas can be obtained from the $K$ lines of CH$_3$CN 
$J=12-11$.  To avoid the simplification of considering optically-thin emission assumed in 
a population-diagram analysis, we fit all the $K$ lines assuming LTE, 
while simultaneously solving for the temperature $T_\mathrm{kin}$, column density of CH$_3$CN molecules 
$N_\mathrm{CH_3CN}$, and line width at half-power FWHM. 
The procedure to obtain the level populations can be found in \cite{Araya05}. 

Figure \ref{fig8} shows the results of our fits to the CH$_3$CN spectra. The systemic  
velocity $V_\mathrm{sys}\approx 38.5~\kms$ was found to be optimal. The data outside 
the lines of interest have been suppressed for clarity, and the fit was done in the frequency 
windows where only the lines of interest are present. 
The gas is warmer (by $18\%$) and denser (by $415\%$ in column) 
toward the peak MM1-Main ({\it bottom} frame)  than in the average 
of the sources composing MM1 ({\it top} frame). This makes the case for the internal heating and a 
centrally-peaked 
density gradient in MM1-Main, as well as for its protostellar nature. 
Some lines are not completely well fit under the assumption of a single value for the parameters. 
On the one hand, the bright $K=7$ and 8 lines indicate the presence of some column of very warm gas; 
on the other hand, the $K=3$ is not brighter than the $K<3$ lines within the uncertainties, 
indicating that some column of gas is below 200 K. Two-component fits do not give better results. 
The reported values should be interpreted as an average 
along the line of sight. Detailed radiative transfer in the context of a 
physical model is currently under way. 

Under the interpretation of edge-on rotation as the cause of the velocity gradient observed in 
Figs. \ref{fig6} and \ref{fig7}, the dynamical mass in 
MM1-Main (stars plus gas) is about $9~\Msun$, where a mean-velocity offset of 
2.0 $\kms$ with respect to $V_\mathrm{sys}\approx38.5$ $\kms$ was taken at a radius of $0.5\arcsec$. 
Assuming that the disk is perpendicular to the outflow, and correcting for the outflow inclination 
angle $i=50^\circ$ estimated by \cite{dW09}, the enclosed dynamical mass is $M_\mathrm{dyn} \sim 15~\Msun$.  

Given that W33A is fragmented into multiple sources, 
a strict upper limit to the stellar mass in MM1-Main is $<20~\Msun$, the mass necessary to account 
for the total luminosity ($\sim 10^5~L_\odot$) of W33A. 
Also, the total gas mass in the 
MM1 sources inferred from dust emission is $\sim 10~\Msun$, therefore the gas mass in the rotating 
structure around MM1-Main should be a fraction of it.  
For a gas mass in the rotating structure $M_\mathrm{gas}\sim5~\Msun$, 
the mass of the protostar(s) in MM1-Main amounts to $M_\star\sim10~\Msun$. 
MM1-Main appears to be B-type protostar still accreting from a rotating disk-like structure. 

From the $\sim [9,32]\Msun$ of gas mass in MM1 we derive an average column density of molecular gas 
of about $[1,4]\times10^{23}$ \column. For an average CH$_3$CN column of $\sim 8\times10^{15}$ \column, 
a CH$_3$CN abundance with respect to H$_2$ of $X$(CH$_3$CN) $\sim [2,8]\times 10^{-8}$ is derived, 
similar estimates in other regions of high-mass star formation \citep{Wil94,Remij04,GM09}.

\section{Discussion} \label{discussion}

\subsection{Star Formation from Converging Filaments}

In W33A, two localized regions of star formation (MM1 and MM2) separated by 0.1 pc are surrounded by a common 
filamentary structure of $\sim 1$ pc in length (Section \ref{parsec}). 
The two velocity components of this filamentary structure are separated by $\approx 2.6~\kms$ in 
line-of-sight velocity and 
intersect in projection right at the position of the star formation activity (Fig. 2). 
The velocity components are not a mere superposition in position--position--velocity space, but they merge into a 
structure with larger motions (Fig. \ref{fig3}, Section ~ \ref{parsec}), resolved into a disk/outflow 
system by subarcsecond resolution observations (Figs. \ref{fig6} and \ref{fig7}, Section ~ \ref{innermost}). 
This suggests that star formation in W33A was triggered by the convergence of  
molecular filaments. Such a scenario has been suggested for the region W3 IRS 5 by 
\cite{Rodon08}. More recently, the merging of filaments has also been claimed by 
\cite{JS10} in the infrared dark cloud G35.39--0.33 and by \cite{Carrasco10} in the MSFR W75.

This mode of star formation is predicted by numerical simulations of star formation triggered by 
converging flows \citep{BP99,Heit08,VS07}. 
In those simulations, the formation of molecular clouds itself is a product of the convergence of streams 
of neutral gas. Later in the evolution of the molecular clouds, filaments of molecular gas can 
converge (merge) with each other, leading to the formation of cores and stars.

We present here a simple comparison with a region found in the numerical
simulation recently reported by \cite{VS09}. This
simulation represents the formation of a giant molecular complex from
the convergence of two streams of warm neutral gas, at the scale of tens
of parsecs. Specifically, the simulation was performed using the
smoothed particle hydrodynamics (SPH) code GADGET \citep{Spring01}, 
including sink-particle and radiative cooling
prescriptions \citep{Jap05,VS07}. The
convergence of the warm diffuse-gas streams  
triggers a thermal instability in the gas, which causes it to undergo a
transition to the cold atomic phase, forming a cloud. The latter soon
becomes gravitationally unstable, begins contracting, and undergoes 
hierarchical fragmentation. During the contraction, the density of
isolated clumps increases and they can reach physical conditions
corresponding to those of molecular clumps. Finally, the global collapse
reaches the center of mass of the cloud, at which point a region with
physical conditions corresponding to those of MSFRs forms. The simulation 
box has 256 pc 
per side, and the converging flows have a length of 112 pc, and a radius
of 32 pc. However, after the gravitational contraction, the clumps
are only a few parsecs across. Since SPH is essentially
a Lagrangian method, it allows sufficient resolution in these dense
regions. We refer the reader to \cite{VS07,VS09} for details of the
simulation. Here we focus on the region called ``Cloud 1'' in \cite{VS09},
albeit roughly 1.5 Myr later than the time examined in that paper.

Figure \ref{fig9} shows two snapshots of column density separated by
0.133 Myr (the time interval between successive data dumps of the
simulation). The column density is computed by integrating the density
along the $x$-direction over the 10 pc path $123 \le x \le 133$ pc,
which is centered at the midplane of the simulation, where the
(sheet-like) cloud is located. In this region and epoch of the
simulation, the two leftmost filaments in the {\it top panel} of Fig.\
\ref{fig9} converge to form the boomerang-shaped filament seen in the
{\it bottom panel}. 
Note that the simulation was
not designed to simulate the observed filaments. 
The comparison is only
intended to show that some of observed properties of W33A can arise
naturally in the context of a simulation of the formation of a large
molecular complex that contains a filament system.

The peaks of two filaments of gas are initially separated by
0.4 pc ({\it top} frame of Fig. \ref{fig9}) and then merge at a projected velocity of
$3\pm1.5$ $\kms$, measured directly from the displacement observed
between the two panels. The column density of the filaments is in the range
[$10^3,10^4$] code units (Fig. \ref{fig9}), or
$N_\mathrm{H_2}=[0.5\times10^{24},0.5\times10^{25}$] cm$^{-2}$. For an NH$_3$
abundance with respect to H$_2$ in the range $X(NH_3)=[10^{-8},10^{-7}]$
\citep[e.g.,][]{GM09}, the column density of the cold
filaments in our observations is
$N_\mathrm{H_2}=[1.1\times10^{24},1.1\times10^{25}$] cm$^{-2}$.  
Figure \ref{fig10} shows the volume density (color scale) and $y$--$z$ plane velocity (arrows)
in a slice trough the filaments. It is seen that the filaments reach densities typical 
of MSFRs (peak $n\sim10^5$ cm$^{-3}$) and that their 
velocity field presents fast jumps of 
a few $\kms$ in the interaction zones, comparable to our observations. 
We conclude that some of the 
properties of the observed filaments such as sizes, column densities, 
and velocities agree within a factor of 2 with those 
from the simulation. This rough comparison illustrates that our
interpretation of convergence between the observed filaments is feasible.

\subsection{Cores at Different Evolutionary Stages}

The star-forming cores in W33A appear to be 
at markedly different evolutionary stages (Section \ref{innermost}). The first piece of evidence for this 
is the clear difference in the richness of the molecular-line emission from MM1 to MM2 
(see Fig. \ref{fig5}). MM1 has molecular emission typical of a ``hot-core'' \citep{Kurtz00}, 
with a prominent CH$_3$CN $J=12-11$ ``$K$-forest'' that can be detected up to the $K=8$ line, 
with upper-level energy $E_u=525.5$ K (see Table 3). The average gas temperature of MM1 is 
$\sim 347$ K. In contrast, MM2 is almost devoid of 
``hot-core'' emission, and is only detected in a few molecules. MM2 is much colder than MM1, with 
a temperature $T_\mathrm{MM2}\approx 46$ K. The second piece of evidence is 
the mass content of the cores. MM1 has only $\sim [9,32]~M_\odot$ of gas, while MM2 
has $\sim 60~\Msun$. This could naively be interpreted as MM2 having a much larger 
gas reservoir than MM1, but it should be kept in mind that both cores are part of a common parsec-scale 
structure. Clump infall at pc scales has been reported from single-dish \citep{Wu03} and 
interferometric \citep{GM09} observations. 
Also, numerical simulations of parsec-sized clumps show that the star-forming cores that give birth 
to massive stars are continuously fed from gas in the environment at the clump scales 
\citep[e.g.,][]{Bonn03,Smith09,VS09,Peters10}. 

There are two possibilities that we briefly discuss here: 
(1) the prestellar cores that were the precursors to MM1 and MM2 appeared 
at the same time and then MM1 evolved faster to produce a $\sim 10~\Msun$ star (Section \ref{innermost}), 
while MM2 
only produced at most an intermediate mass star (MM2 is not prestellar, since it has some 
internal heating and appears to power an outflow, see Section \ref{parsec}), 
or (2) the prestellar core precursor to MM2 formed later 
and has yet to form at least one massive star. 
Our observations cannot tell these options apart. A measurement of the accretion rate 
in both cores  would be helpful. 
Sources at different evolutionary stages within a single star-forming cluster have also 
been reported recently for AFGL 5142 \citep{Zhang07}, G28.34+0.06 \citep{Zhang09}, and 
AFGL 961 \citep{Will09}.

\subsection{A Rotating Disk/Outflow System in MM1-Main}

In the past decade, the question of whether massive stars form by disk-outflow mediated 
accretion similar to low-mass stars has been the subject of intensive research. The answer 
is positive: they definitely do. 
Some of the massive protostars that have been shown to harbor disk/outflow systems 
are G192.16-3.82 \citep{Shep01},  
Cepheus A HW2 \citep{Patel05}, IRAS 20126+4104 \citep{Cesa05}, and 
IRAS 16547-4247 \citep{FH09}. 
All these relatively clean disk examples, however, do not have stars more massive than 15 to $20~\Msun$.  
More massive (O-type) stars have also been shown to form via disk-mediated accretion. 
However, the innermost part of the accretion flow is often (at least partially) ionized  
and is observed 
as a hypercompact (HC) \HII region. Also, the gas in these more massive regions is warmed up to farther 
distances, and very massive rotating structures of size up to 0.1 pc are typically observed. 
Examples are G10.6-0.4 \citep{KW06}, G24.78+0.08 \citep{Bel06}, 
W51e2 \citep{Klaass09}, and G20.08-0.14 N \citep{GM09}, all of which have stellar masses above 
$20~M_\odot$.   
A possible exception to this scenario is W51 North, where \cite{Zap08,Zap09} claims 
to have found a protostar with $M_\star > 60~\Msun$ and without a ``bright'' (with flux above 
tens of mJy at wavelengths $\sim 1$ cm) \HII region. This apparent discrepancy is solved 
if multiple, lower-mass stars account for the mass in W51 North or if the \HII region 
in this source is gravitationally trapped as currently observed. Indeed, detailed simulations 
of the evolution of HC \HII regions within accretion flows show that their 
radio-continuum emission flickers significantly in timescales from 10 to $10^4$ yr  
\citep{Peters10}. 

In this paper we report on the existence of a rotating disk centered on MM1-Main in 
W33A (Section \ref{innermost}). To our sensitivity, the radius of the disk is $R \lesssim 4000$ AU. 
The warm-gas emission does not come only from the disk, but also from an structure  
elongated perpendicular to it, coincident with the other mm peaks (MM1-NW and MM1-SE) 
along the direction of the outflow (see Figs. \ref{fig6} 
and \ref{fig7}). We propose 
that the secondary mm peaks in MM1 are not of protostellar nature, but regions where 
the emission is enhanced due to the interaction of the outflow with the disk and its inner 
envelope. 
Indeed, the two brightest 7-mm detections of \cite{vdTM05} are counterparts of 
MM1-Main (Q1, $S_\nu\approx1.7$ mJy at 7 mm) and MM1-SE (Q2, $S_\nu\approx0.6$ mJy at 7 mm). 

The free--free emission from the 7 mm source Q1 should be a combination of 
photoionization by the central 
protostar and shock-induced ionization of material due to the 
jet observed by \cite{Davies10}, likely 
dominated by the latter. The origin 
of the free--free emission from Q2 could also be shocks, although deeper 7 mm observations 
are needed to confirm this source. Q3 does not have a mm/submm counterpart and may not be 
a real source. The radio continuum sources of 
W33A could then be analogs of those in IRAS 16547--4247 \citep{Rod05,Rod08}, 
which have fluxes a factor of a few larger at 0.75 times the distance to W33A. 
Indeed, we find that the radio-continuum emission from Q1 (MM1-Main) agrees with the correlation 
found for low-mass jets between the radio-continuum luminosity of the jet and the momentum 
rate of the associated molecular outflow: $\dot{P}=10^{-2.5}(S_\nu d^2)^{1.1}$ \citep{Angla98}, 
where the radio luminosity 
has units of mJy kpc$^2$ and the momentum rate has units of $M_\odot$ yr$^{-1}$ km s$^{-1}$. 
For W33A with a radio flux  $S_\mathrm{3.6cm}=0.79$ mJy \citep{RH96}, and a distance $d=3.8$ kpc \citep{Jaffe82}, 
the expected momentum rate is $\dot{P}=0.046~\Msun$ yr$^{-1}~\kms$, while the observed momentum rate 
(lower limit) is $\dot{P}=0.040~\Msun$ yr$^{-1}~\kms$ (obtained dividing the momentum 
of the molecular outflow by its length, including the inclination correction). 
This correlation was found to hold for three well-studied massive protostars (IRAS 
16547--4247, HH 80--81, and Cep A HW2) by 
\cite{Rod08}. In this paper we report that it also holds for W33A, which constitutes further 
evidence for a common accretion mechanism between low- and high-mass protostars, at least to 
the stage prior to the development of a brighter \HII region. 

Comparing the ratio of the radio luminosity to the IR luminosity $L$(8 GHz)/$L$(IR) 
with the recombination-line line width 
may be a useful criterion to distinguish between a source 
ionized by shocks (jet or stellar wind) or by photoionization (what usually is called an \HII 
region). In Fig. 6 of the review by \cite{Hoare07}, it is seen that UC \HII regions have 
the largest $L$(8 GHz)/$L$(IR) and the smallest line width, jet sources have the smallest 
$L$(8 GHz)/$L$(IR) and the largest line width, and HC \HII regions fall in between the 
previous two. For W33A, $\log ( L$(8 GHz)/$L$(IR)) $\approx 7.1$, and the FWHM of the IR recombination 
lines is of several hundreds $\kms$, again consistent with a jet source.

The star(s) at the center of MM1-Main (with $M_\star\sim 10~\Msun$) 
appears to dominate the dynamics of the disk, but  
we cannot rule out the existence of additional, less massive objects within it. 
Indeed, models of massive protostellar disks predict their fragmentation 
and the formation of a few lower mass companions within it \citep{KraMa09,Krum09,Peters10}.

\section{Summary and Conclusions} \label{conclu}

We present for the first time resolved observations in both mm continuum and molecular-line 
emission for the massive star formation region W33A, characterized by a very high luminosity 
($L\sim10^5~L_\odot$) and very low radio-continuum emission ($\sim 1$ mJy). Both of the previously 
known mm cores (MM1 and MM2) are resolved into multiple peaks, and appear to be at very different evolutionary 
stages, as indicated by their molecular spectra, masses, temperatures, and continuum spectral 
indices. The brightest core (MM1-Main at the center of MM1) is centered on a very faint 
free--free source and the gas dynamics up to a few thousand AU of it indicates the presence of a 
circumstellar disk rotating around a stellar mass of $M_\star \sim 10~\Msun$. 
MM1-Main also drives a powerful, high-velocity molecular outflow perpendicular to the disk. 
MM2, the coldest and most massive core, is not detected in hot-core lines but appears to 
drive a more modest outflow. Both MM1 and MM2 are located at the intersection of 
parsec-scale filamentary structures with line-of-sight velocity offset by $\approx 2.6~\kms$. 
Analysis of the position--position--velocity structure of these filaments and a comparison with 
recent numerical simulations 
suggests that star formation in W33A was triggered by the convergence of filaments of cold molecular gas.

\acknowledgements

R.G.M. acknowledges support from SAO and ASIAA  through an SMA predoctoral fellowship. 
J.E.P. is supported by the NSF through grant AF002 from the Association of Universities for Research in 
Astronomy, Inc., under NSF cooperative agreement AST-9613615 and by Fundaci\'on Andes under project 
no. C-13442. Support for this work was provided by the NSF through awards GSSP06-0015 and GSSP08-0031 
from the NRAO. 

We thank Ben Davies for his clarifications regarding his published results and 
the anonymous referee for comments that helped us to improve this paper.

\clearpage

\begin{center} 
\begin{deluxetable}{ccccccc}  \label{tab1}
\tabletypesize{\scriptsize}
\tablecaption{Observational Parameters}
\tablewidth{0pt}
\tablehead{
\colhead{Epoch} & \colhead{Array} & \multicolumn{2}{c}{Phase Center\tablenotemark{a}} & 
\colhead{Bandpass} & \colhead{Flux} & \colhead{Phase} \\\cline{3-4} 
 &  &  $\alpha$(J2000) & $\delta$(J2000) & Calibrator & Calibrator & Calibrator 
}
\startdata
2004 06 14+15 & VLA-D & 18 14 39.500 & $-17$ 51 59.800 & 3C273 & 3C286 & $1851+005$  \\
2006 05 22 & SMA-Extended & 18 14 39.509 & $-17$ 51 59.999 & 3C273 & Callisto & $1733-130$ \\
2007 07 17 & SMA-Compact & 18 14 39.495 & $-17$ 51 59.800 & 3C454.3 & Callisto & $1733-130$ \\
2008 08 02 & SMA-VEX & 18 14 39.495 & $-17$ 51 59.800 & 3C454.3 & Callisto & $1733-130$ \\
\hline
\enddata
\tablenotetext{a}{Units of R.A. are hours, minutes, and seconds. Units of decl. are 
degrees, arcminutes, and arcseconds.}
\end{deluxetable}
\end{center}

\begin{center} 
\begin{deluxetable}{cccccc}  \label{tab2}
\tabletypesize{\scriptsize}
\tablecaption{Millimeter Continuum Sources}
\tablewidth{0pt}
\tablehead{
\colhead{Core\tablenotemark{a}} & \colhead{Component\tablenotemark{b}} & 
\colhead{$\alpha$(J2000)\tablenotemark{c}} & 
\colhead{$\delta$(J2000)\tablenotemark{c}} & 
\colhead{$I_\mathrm{peak}$(1.3 mm)\tablenotemark{d}} & 
\colhead{$S$(1.3 mm)\tablenotemark{e}} \\
 &  & (hrs,min,sec) & (deg,arcmin,arcsec) & ($\mjyb$)  & (mJy)
}
\startdata
    & MM1-NW   & 18 14 39.47 & $-17$ 51 59.7 & 31 & 357   \\ 
MM1 & MM1-Main & 18 14 39.51 & $-17$ 52 00.0 & 65 & 357  \\ 
    & MM1-SE   & 18 14 39.55 & $-17$ 52 00.4 & 25 & 357 \\
\hline
MM2 & MM2-Main & 18 14 39.24  & $-17$ 52 01.9 & 43  & 289 \\
    & MM2-NE   & 18 14 39.31  & $-17$ 52 00.6 & 22 & 289 \\
\hline
\enddata
\tablenotetext{a}{Main core as labeled in Fig. \ref{fig1}, left panel.}
\tablenotetext{b}{Clearly distinct subcomponents of the main cores as marked in 
Fig. \ref{fig1}.} 
\tablenotetext{c}{Position of peak.}
\tablenotetext{d}{Peak intensity $\pm 1.5$ $\mjyb$. HPBW $=0.63\arcsec \times 0.43\arcsec$.}
\tablenotetext{e}{Added flux of the subcomponents of each core. The uncertainties in the fluxes of 
MM1 and MM2 are $\pm20$ mJy and $\pm25$ mJy respectively.}

\end{deluxetable}
\end{center}

\clearpage

\begin{center} 
\begin{deluxetable}{cccc}  \label{tab3}
\tabletypesize{\scriptsize}
\tablecaption{Bright Molecular Lines\tablenotemark{a}}
\tablewidth{0pt}
\tablehead{
\colhead{Species} & \colhead{Transition} & \colhead{$\nu_0$}  & \colhead{$E_u$} \\
   &   & (GHz)  & (K)  \\
}   
\startdata
C$^{18}$O & 2--1 & 219.5603  & 15.8  \\
HNCO & 10(2,9)--9(2,8) & 219.7338  & 228.4 \\
-- & 10(2,8)--9(2,7) & 219.7371  & 228.2 \\
-- & 10(0,10)--9(0,9) & 219.7982  & 58.0 \\
H$_2$$^{13}$CO & 3(1,2)--2(1,1) & 219.9084 & 32.9 \\
SO & 6(5)--5(4) & 219.9494  & 34.9 \\
CH$_3$OH & 8(0,8)--7(1,6) $E$ & 220.0784 & 96.6  \\
$^{13}$CO & 2--1 & 220.3986 & 15.86 \\
CH$_3$CN & 12(8)-11(8) & 220.4758  & 525.5  \\
-- & 12(7)-11(7) & 220.5393  & 418.6 \\
HNCO & 10(1,9)--9(1,8) & 220.5847 &  101.5  \\
CH$_3$CN & 12(6)-11(6) & 220.5944  & 325.8  \\
-- & 12(5)-11(5) & 230.6410 & 247.3 \\
-- & 12(4)-11(4) & 220.6792  & 183.1  \\
-- & 12(3)-11(3) & 220.7090  & 133.1  \\
-- & 12(2)-11(2) & 220.7302  & 97.4 \\
-- & 12(1)-11(1) & 220.7430  & 76.0  \\
-- & 12(0)-11(0) & 220.7472  & 68.8  \\
CH$_3$OH & 15(4,11)--16(3,13) $E$ & 229.5890 & 374.4  \\
-- & 8(-1,8)--7(0,7) $E$ & 229.7588  & 89.1  \\
-- & 3(-2,2)--4(-1,4) $E$ & 230.0270  & 39.8  \\
$^{12}$CO & 2--1 & 230.5380  & 16.5  \\
OCS & 19--18 & 231.0609  & 110.8  \\
$^{13}$CS & 5--4 & 231.2207  & 26.6  \\
CH$_3$OH & 10(2,9)--9(3,6) $A-$ & 231.2811 & 165.3  \\
\enddata

\tablenotetext{a}{Molecular lines with peak $T_B\geq20$ K. The first column refers to the molecule tag, 
the second column to the transition, 
the third to its rest frequency as found in Splatalogue (http://www.splatalogue.net/), and the fourth 
to the upper-level energy.  Data used from Splatalogue are compiled from the CDMS catalog 
\citep{Muller05} and the NIST catalog \citep{Lovas04}}.
\end{deluxetable}
\end{center}

\clearpage




\clearpage 

\clearpage 

\begin{figure}
\figurenum{2b (continued)}
\begin{center}
\includegraphics[angle=0,scale=0.3]{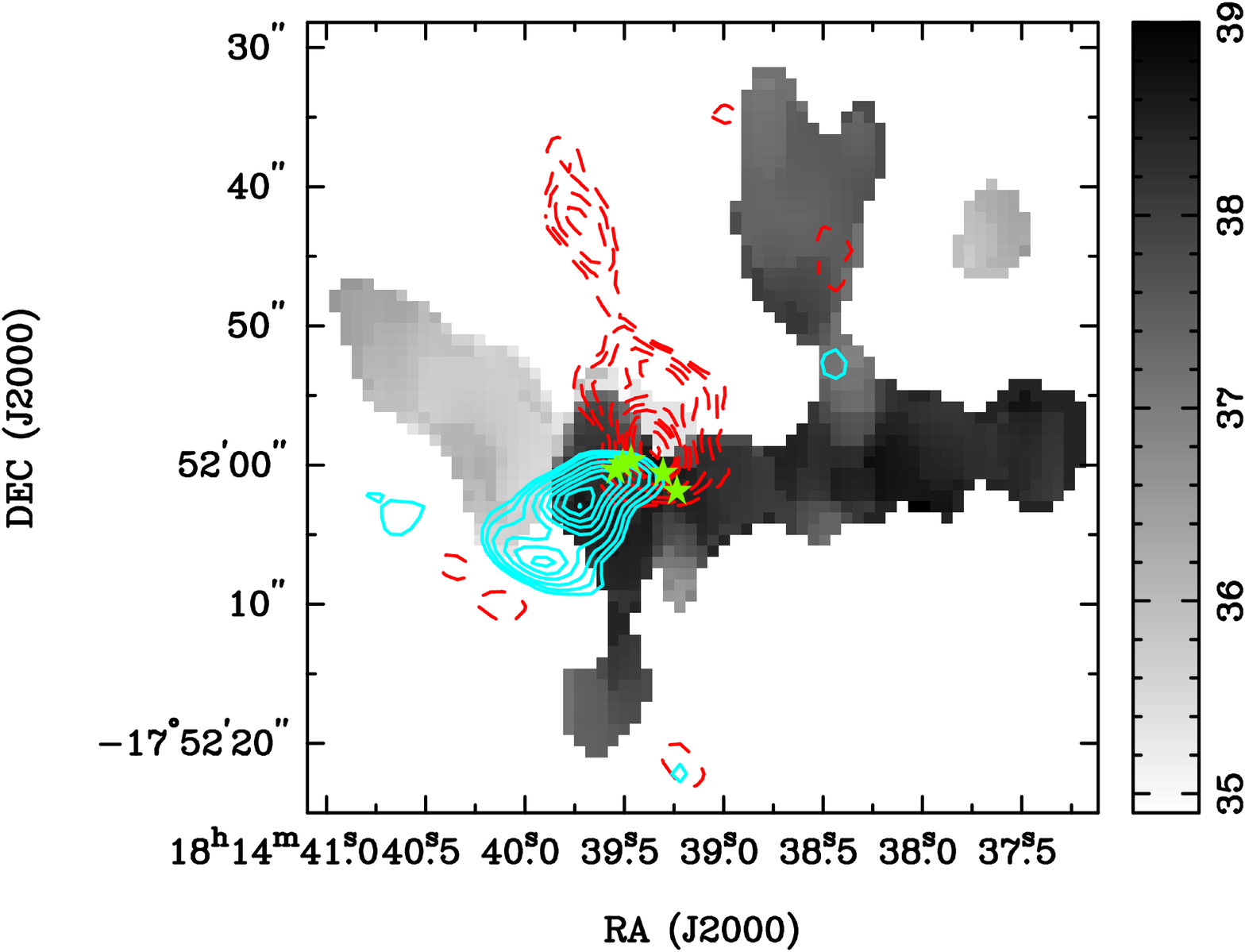}
\end{center}
\caption{
Parsec-scale gas structure toward W33A. The gray scale shows the 
intensity-weighted mean velocity (moment 1, in $\kms$ with respect to the LSR) of the 
quiescent  NH$_3$ (2,2) gas (HPBW $=6.0\arcsec\times2.6\arcsec$, 
P.A. $=2^\circ$). The blue solid contours show the high-velocity gas as detected in CO (2--1) 
(HPBW $=3.0\arcsec\times2.0\arcsec$, P.A. $=56^\circ$) integrated in the range [0,22] $\kms$. 
The red dashed contours show the high-velocity CO gas in the range [62,98] $\kms$. 
Contour levels are $-5,5,7,10,15,20,25,30,35,40,50,$ and $60\times0.7$ $\jybkms$. 
The mm continuum 
sources identified in this paper are marked with stars. 
10 arcsec corresponds to 38,000 AU (0.184 pc).
}
\label{fig2b}
\end{figure}

\clearpage 

\begin{figure}
\figurenum{2c (continued)}
\begin{center}
\includegraphics[angle=0,scale=0.3]{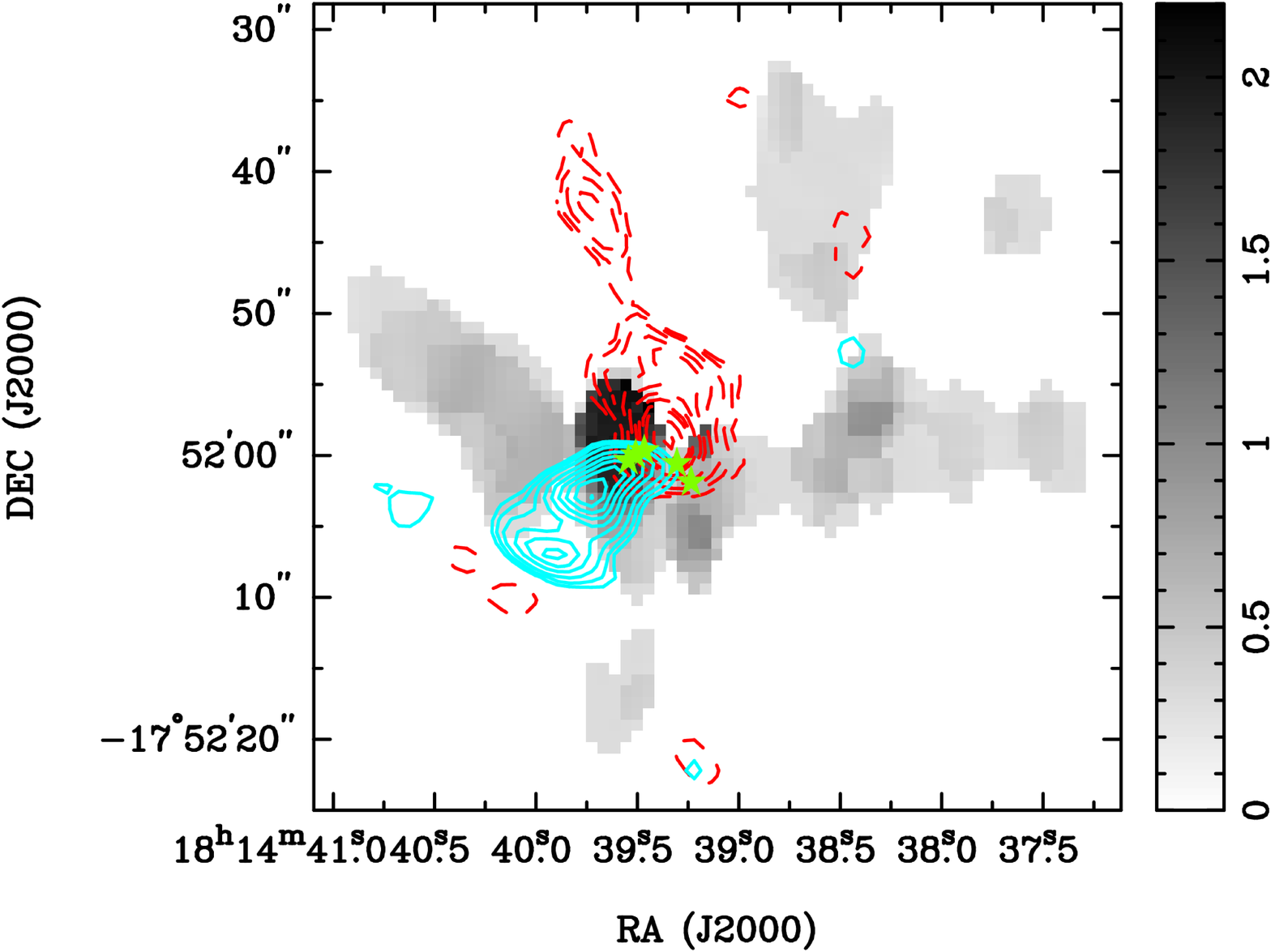}
\end{center}
\caption{
Parsec-scale gas structure toward W33A. The gray scale shows the 
velocity dispersion with respect to the mean velocity (moment 2, FWHM/2.35 in $\kms$)
of the quiescent  NH$_3$ (2,2) gas (HPBW $=6.0\arcsec\times2.6\arcsec$, 
P.A. $=2^\circ$). The blue solid contours show the high-velocity gas as detected in CO (2--1) 
(HPBW $=3.0\arcsec\times2.0\arcsec$, P.A. $=56^\circ$) integrated in the range [0,22] $\kms$. 
The red dashed contours show the high-velocity CO gas in the range [62,98] $\kms$. 
Contour levels are $-5,5,7,10,15,20,25,30,35,40,50,$ and $60\times0.7$ $\jybkms$. 
The mm continuum 
sources identified in this paper are marked with stars. 
10 arcsec corresponds to 38,000 AU (0.184 pc).
}
\label{fig2c}
\end{figure}

\begin{figure}
\figurenum{3}
\begin{center}
\includegraphics[angle=0,scale=0.55]{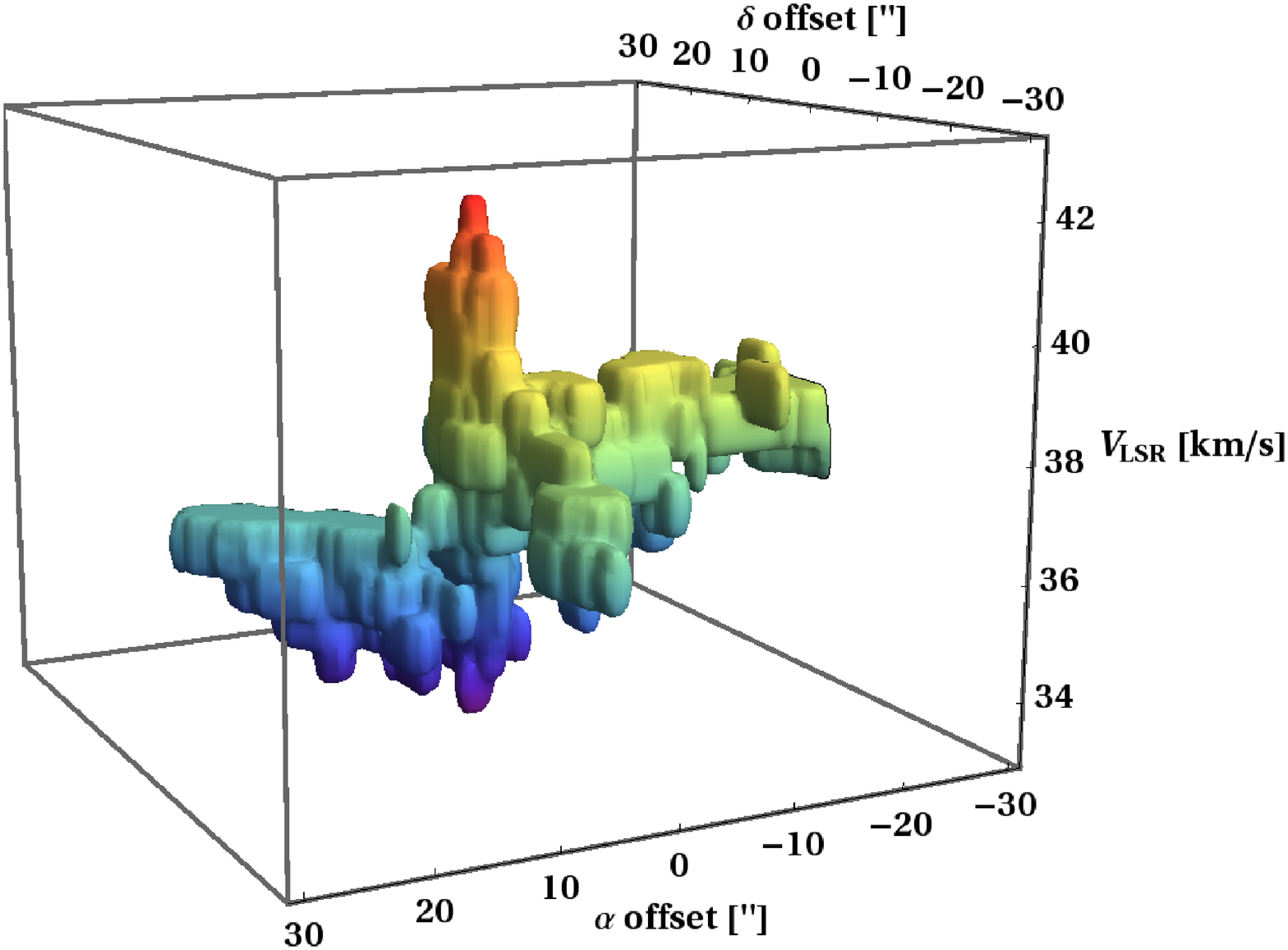}
\end{center}
\caption{
Three-dimensional (position--position--velocity) rendering of the NH$_3$ (2,2) data. Every voxel in the data 
cube with intensity $>20$ $\mjyb$ ($5\sigma$) has been included. The vertical axis is 
color-coded according to $V_{{\rm LSR}}$. It is seen that the two filamentary 
structures at different velocities do not merely superpose at the position center, but merge 
in position--position--velocity space, suggesting interaction of the filaments. 
}
\label{fig3}
\end{figure}

\clearpage

\begin{figure}
\figurenum{4}
\begin{center}
\includegraphics[angle=0,scale=0.8]{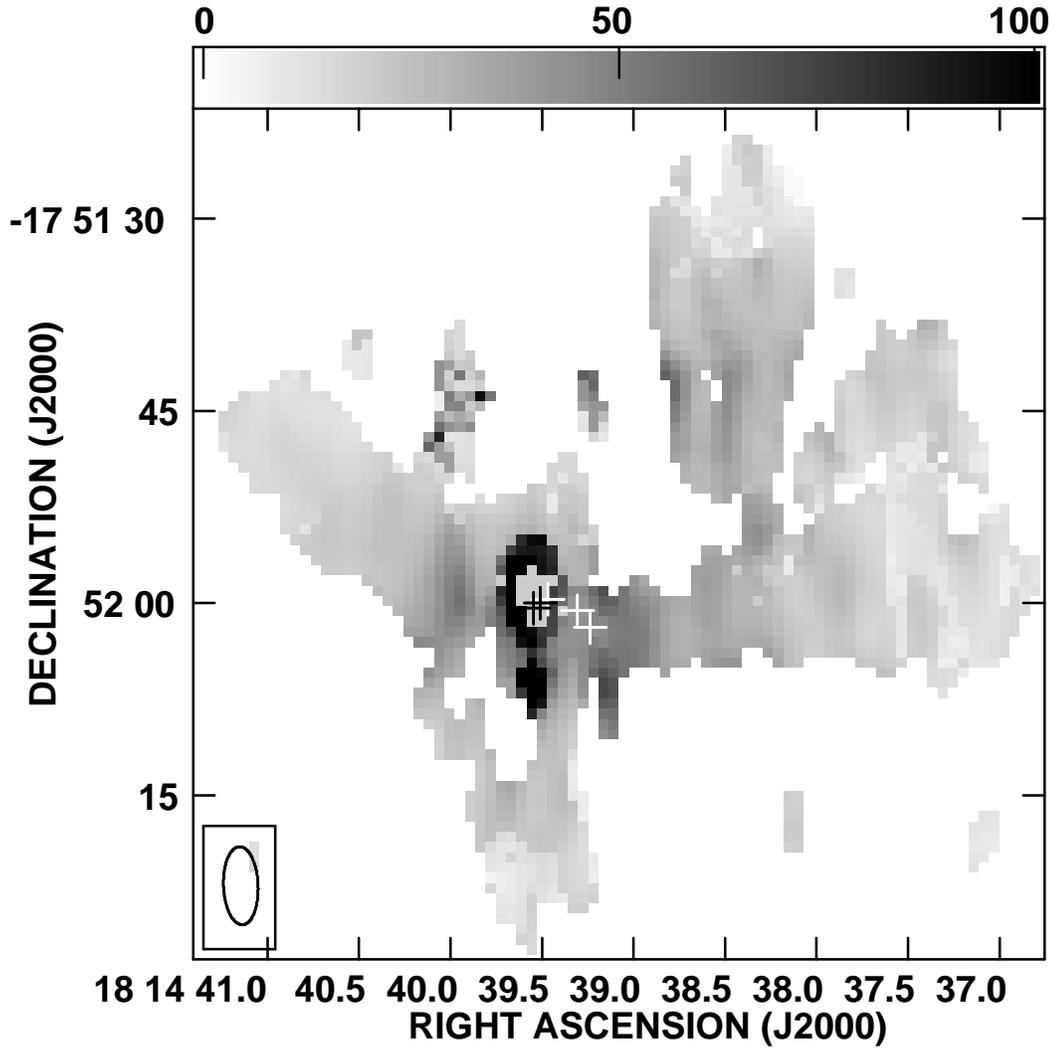}
\end{center}
\caption{
Map of the kinetic temperature $T_\mathrm{kin}$ at large scales obtained from the 
NH$_3$ (1,1) and (2,2) data. 
It is seen that the pc-scale filaments are cold, with $T_\mathrm{kin}=[20,50]$ K. 
Only toward MM1 $T_\mathrm{kin}$ rises significantly, but the errors toward this region 
increase up to $\approx 40$ K. Symbols are as in Fig \ref{fig1}. 
}
\label{fig4}
\end{figure}

\clearpage

\begin{figure}
\figurenum{5}
\includegraphics[angle=0,scale=0.4]{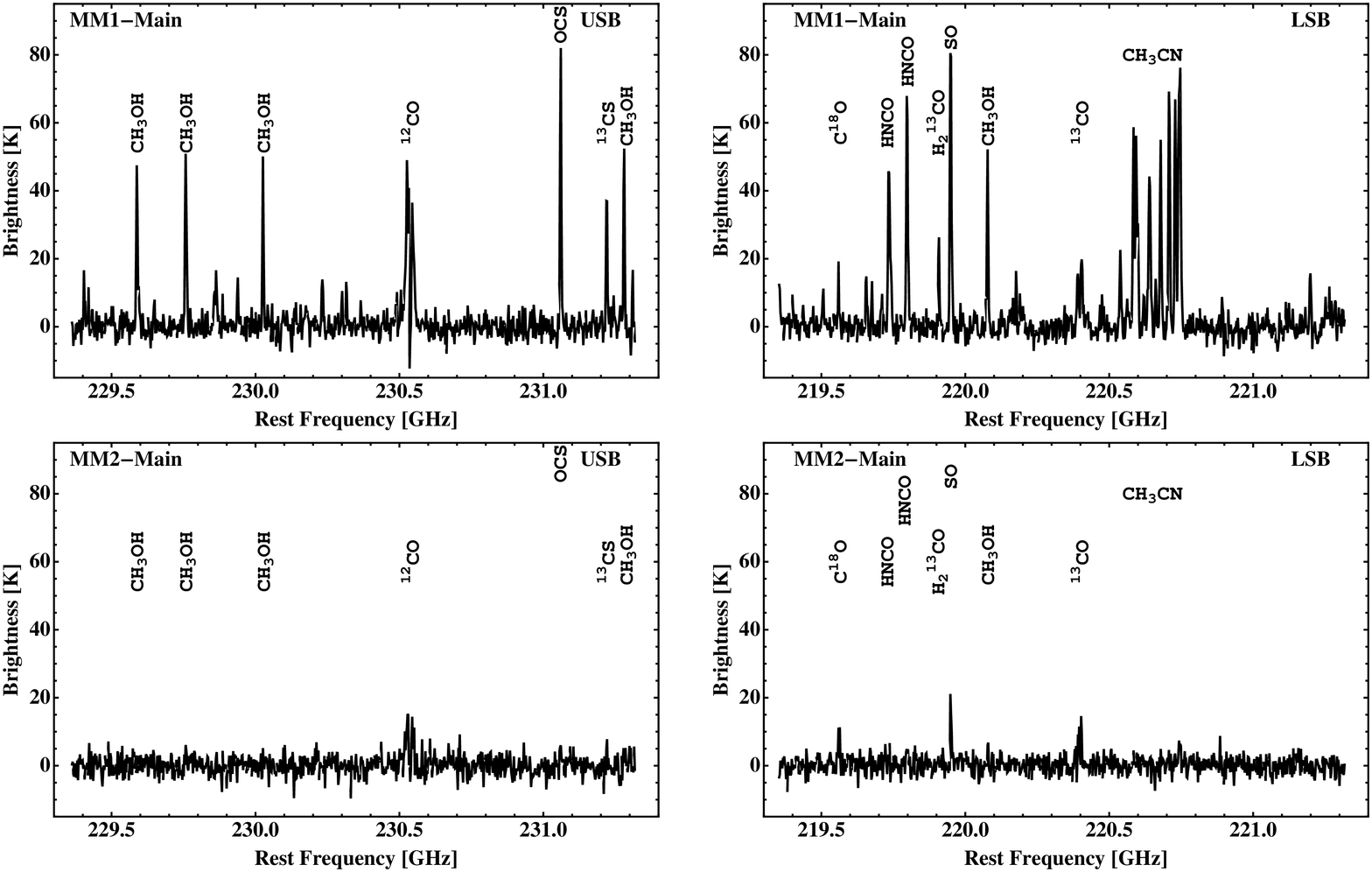}
\caption{
Wide-band, continuum-free spectra in the image domain from the SMA Compact+VEX data at 1.3 mm. 
The {\it top} row shows the spectra for MM1 (the peak MM1-Main). The {\it bottom} row shows 
the spectra for MM2 (the peak MM2-Main). 
There is a striking difference in the richness of the spectra between the two cores. 
MM2 is almost devoid of molecular-line emission, in spite 
of it having a larger gas reservoir than MM1 (Section \ref{sed}). 
}
\label{fig5}
\end{figure}

\clearpage

\begin{figure}
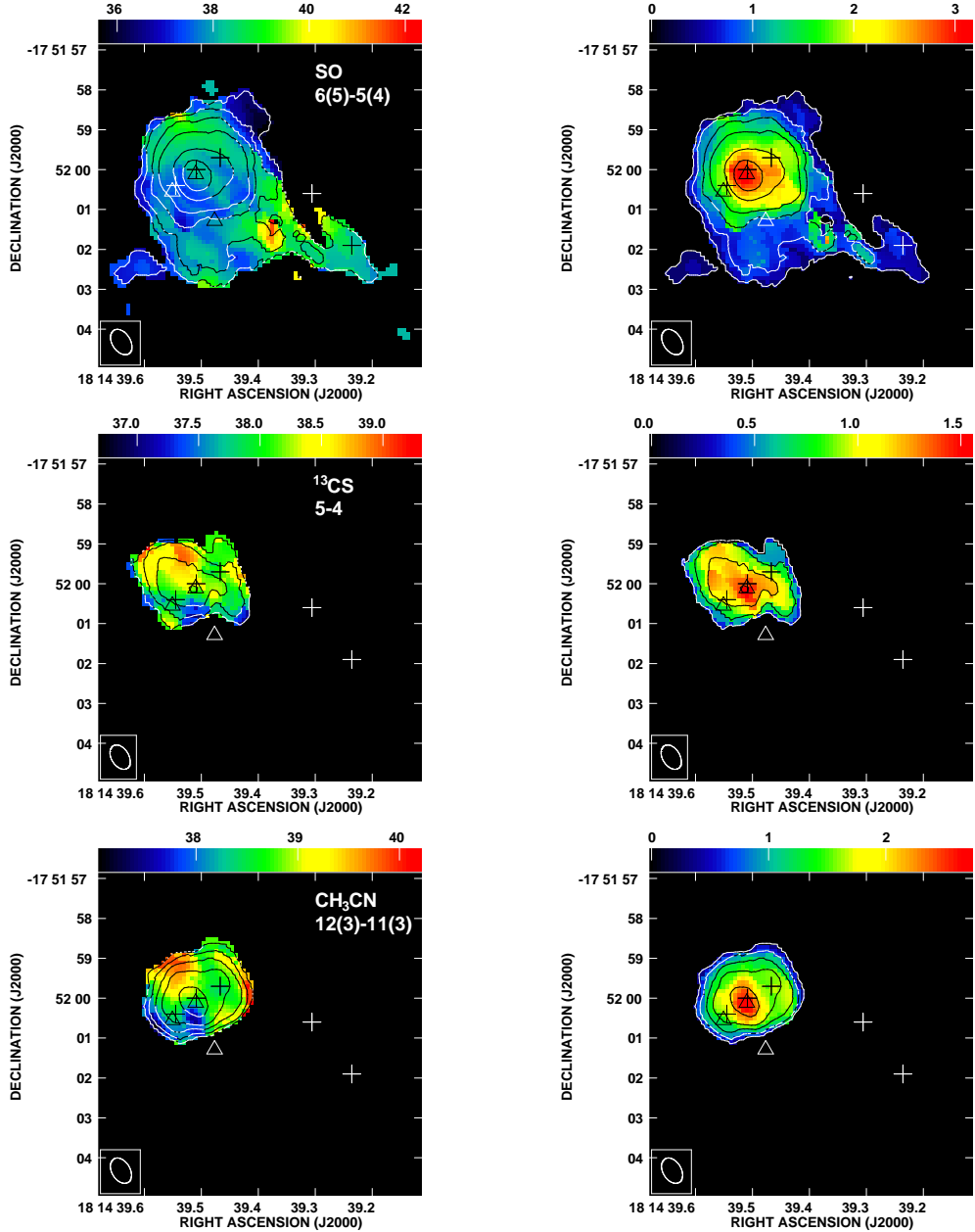

\figurenum{6}
\begin{center}
\epsscale{0.8}
\plottwo{f6a}{f6b}
\plottwo{f6c}{f6d}
\plottwo{f6e}{f6f}
\end{center}
\caption{
Hot-core molecules toward the center of W33A. The {\it top} row shows the SO 6(5)--5(4) line. 
The {\it middle} row shows the $^{13}$CS 5--4 line. The {\it bottom} row shows the CH$_3$CN 
12(3)--11(3) line. Contours show the moment 0 maps at $5,15,30,50,100,150,$ and $200\times0.05$ 
$\jybkms$. The color scale shows the moment 1 maps ({\it left} column) and moment 2 maps 
({\it right} column). Symbols are as in Fig. 1.  While the SO traces an extended envelope reaching 
MM2, the other molecules trace a clear velocity gradient indicative of rotation centered in 
MM1-Main. The velocity dispersion also peaks in MM1-Main. 
}
\label{fig6}
\end{figure}

\clearpage 

\begin{figure}
\figurenum{7}
\begin{center}
\includegraphics[angle=0,scale=0.8]{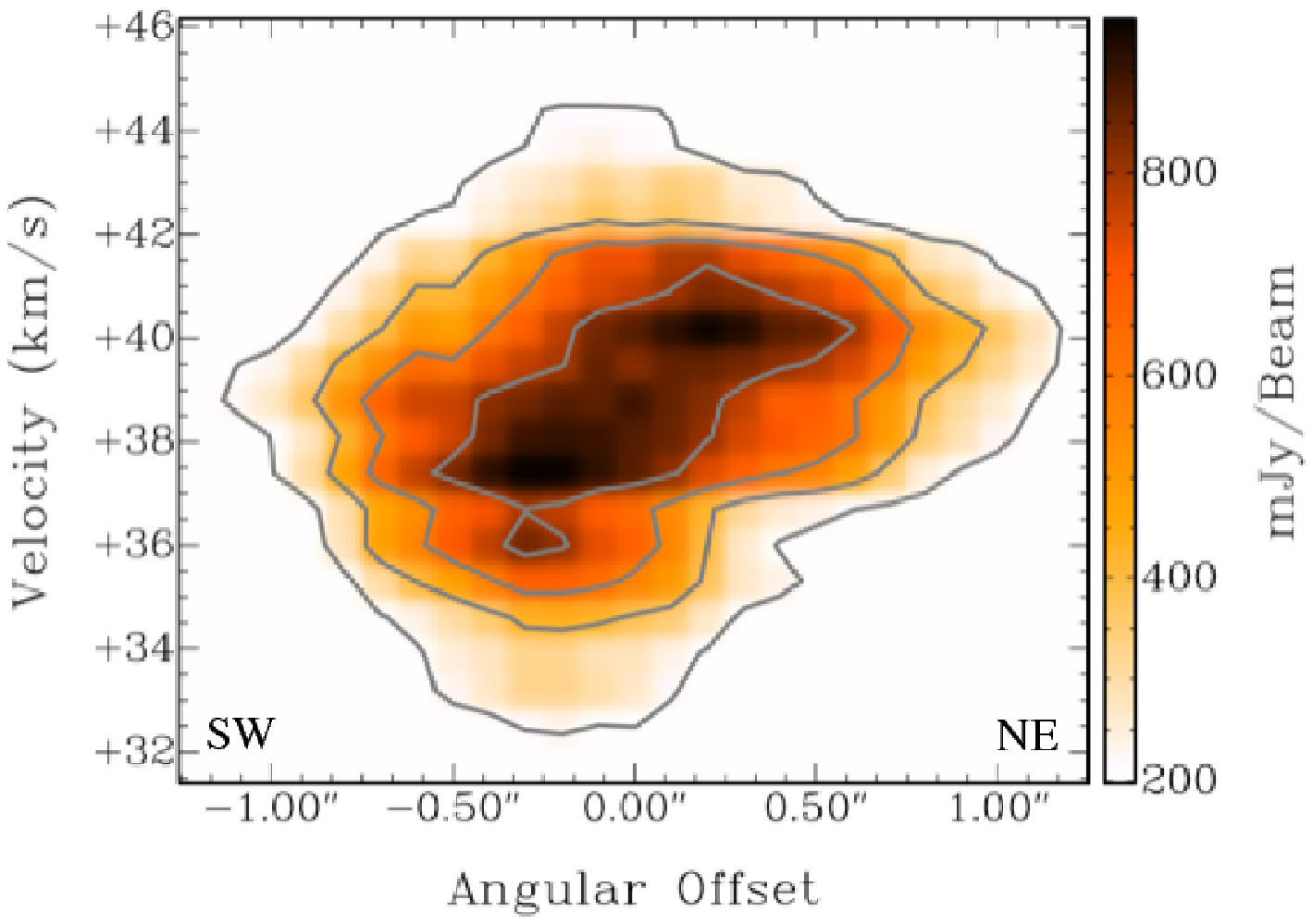}
\includegraphics[angle=0,scale=0.8]{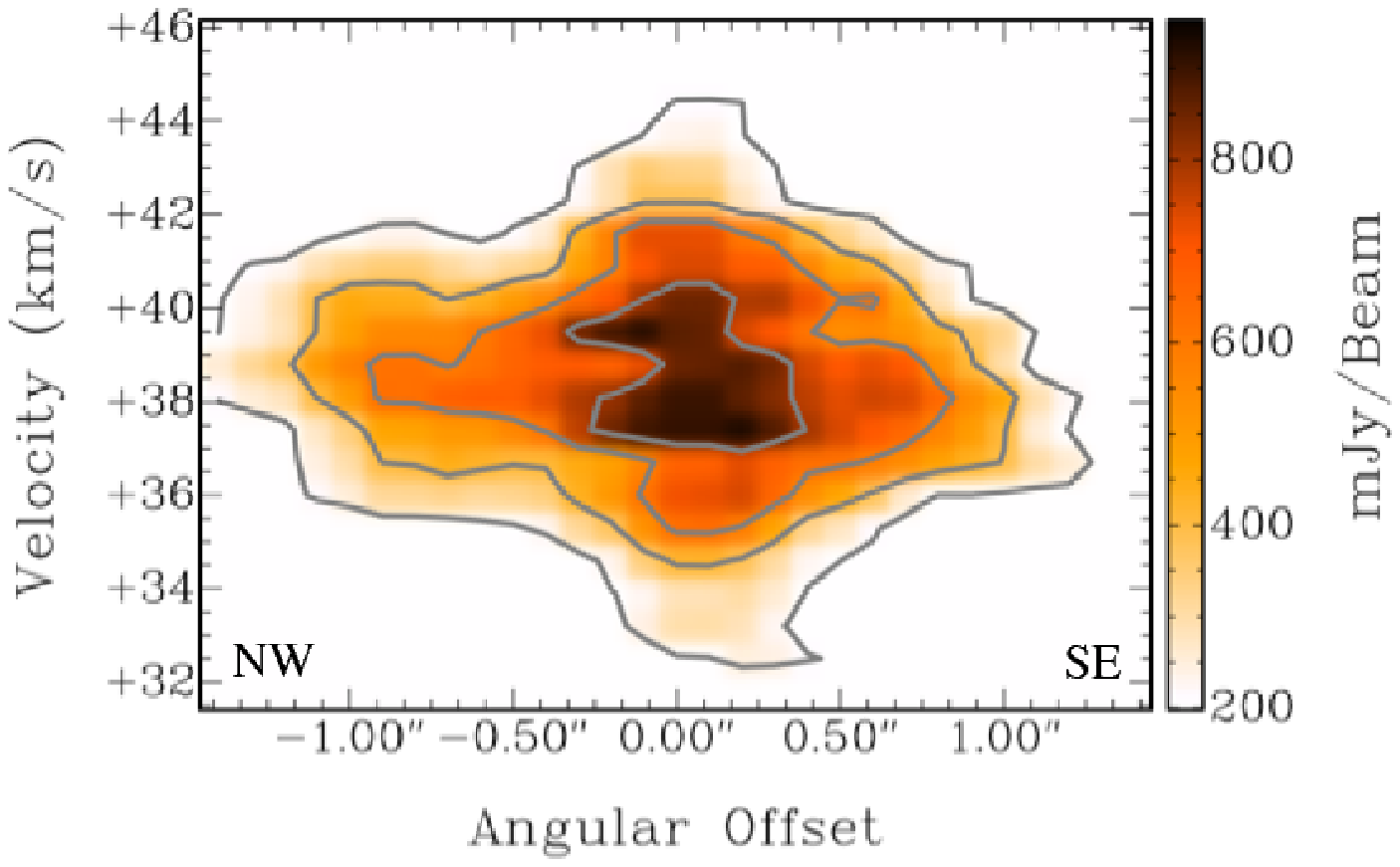}
\end{center}
\caption{
Position--Velocity (PV) diagrams for the CH$_3$CN $J=12-11$ $K=3$ data. The center is the position 
of MM1-Main (Table 2).  {\it Top:} PV diagram at P.A. $=39^\circ$. A clear velocity gradient 
is seen from the southwest (blueshifted, negative position) to the northeast (redshifted, 
positive position). At the angle of this cut the 
velocity gradient has maximum symmetry. {\it Bottom:} PV diagram at P.A. $=39^\circ+90^\circ=129^\circ$. 
Negative positions are 
to the northwest, and positive positions to the southeast. There is emission in the range 
$[36,41]$ $\kms$ at all positions. Closer to the position center, the velocity dispersion 
increases rapidly. 
}
\label{fig7}
\end{figure}

\clearpage 

\begin{figure}
\figurenum{8}
\begin{center}
\includegraphics[angle=0,scale=0.6]{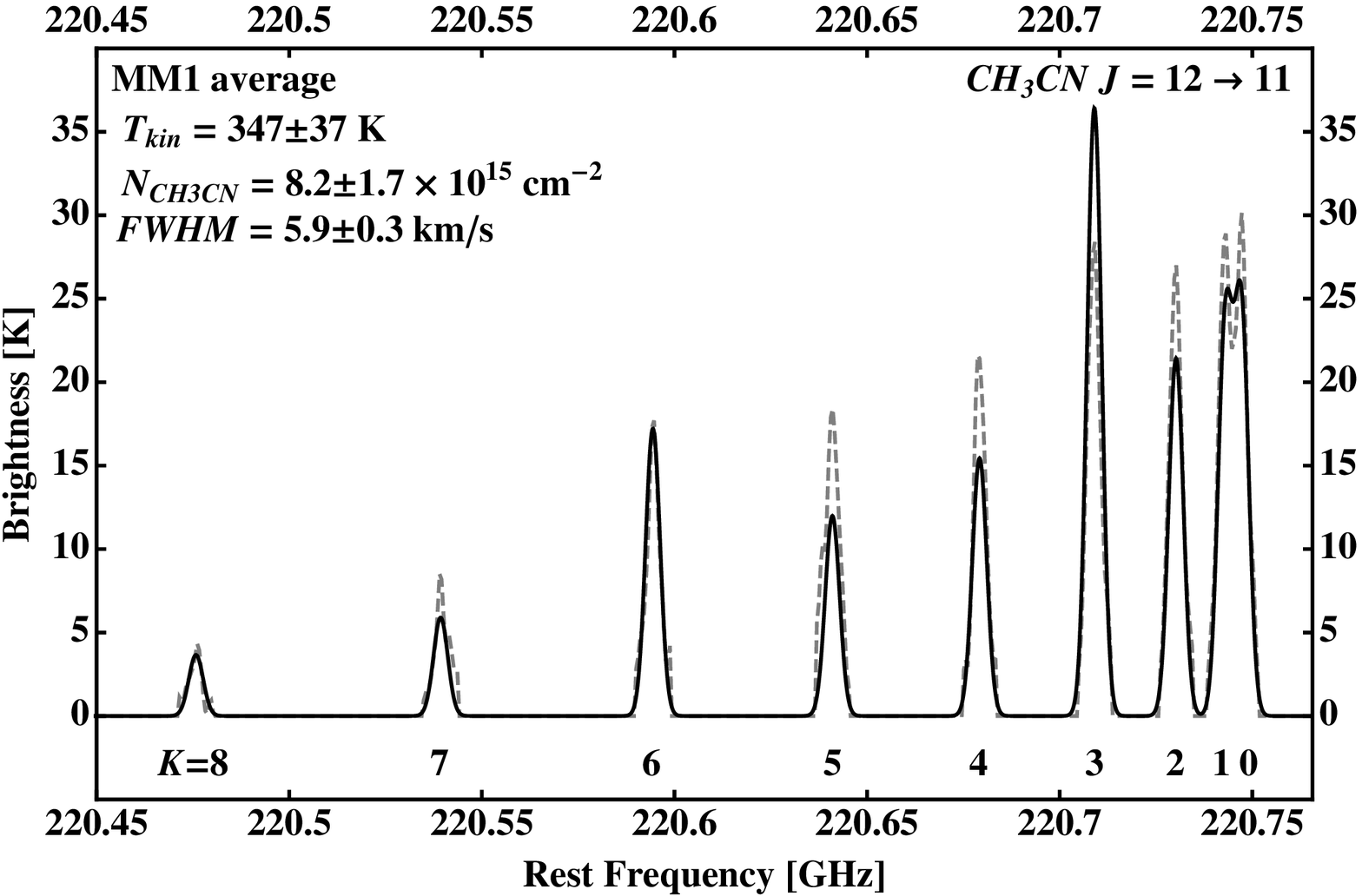}
\includegraphics[angle=0,scale=0.6]{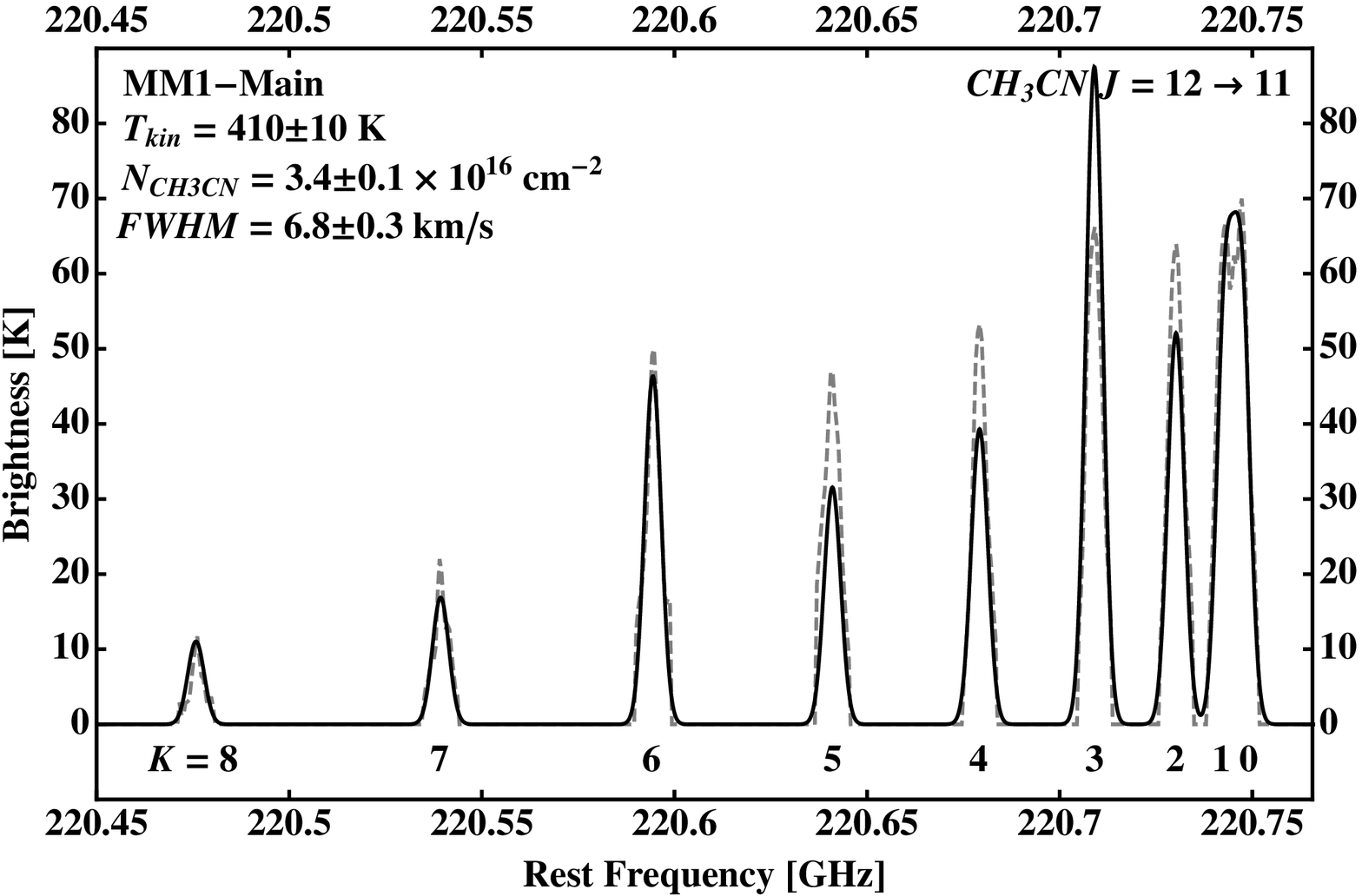}
\end{center}
\caption{
Simultaneous fits to the CH$_3$CN $J=12-11$ $K$ lines.  The dashed gray line is the 
data. The solid black line is the fit. The data outside the lines of interest have  
been set to zero to avoid contamination by other molecular lines. 
{\it Top:} Average spectra over the 
entire MM1 region. {\it Bottom:} Spectra toward the few central pixels at the peak position (MM1-Main). 
The gas is warmer, denser, and has a larger line width toward this position. 
}
\label{fig8}
\end{figure}

\clearpage

\begin{figure}
\figurenum{9}
\begin{center}
\includegraphics[angle=180,scale=0.45]{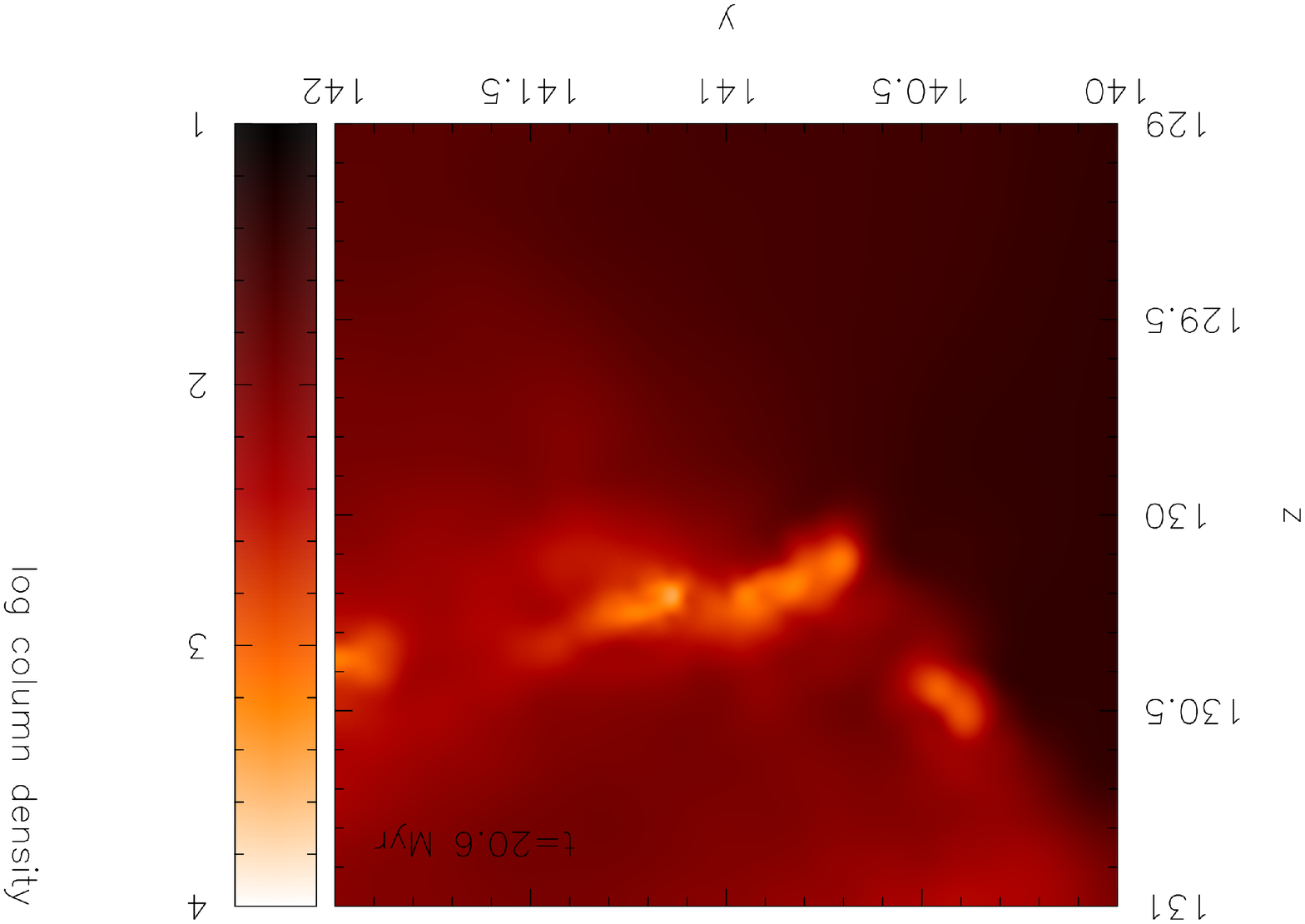}
\includegraphics[angle=180,scale=0.45]{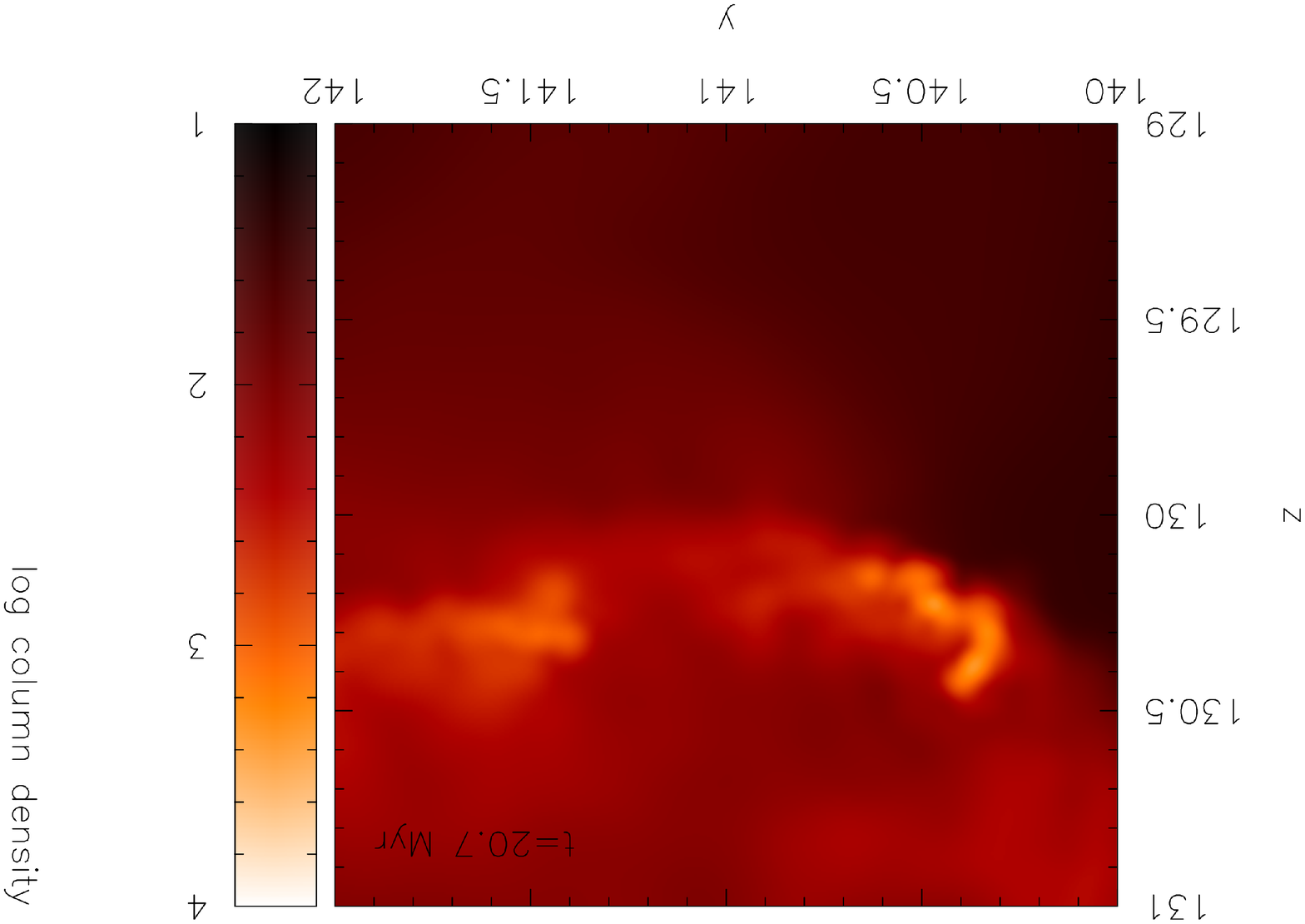}
\end{center}
\caption{
Snapshots of molecular filaments merging with each other extracted from the simulation presented 
by V\'azquez-Semadeni et al. (2009). The time interval between the first ({\it top}) and second 
({\it bottom}) frames is 0.133 Myr. The units of the $z$- and $y$-axes are pc. 
The color scale shows the column density in code units, equivalent to $4.9\times10^{20}$ H$_2$ 
particles per square cm. 
}
\label{fig9}
\end{figure}

\begin{figure}
\figurenum{10}
\begin{center}
\includegraphics[angle=0,scale=0.45]{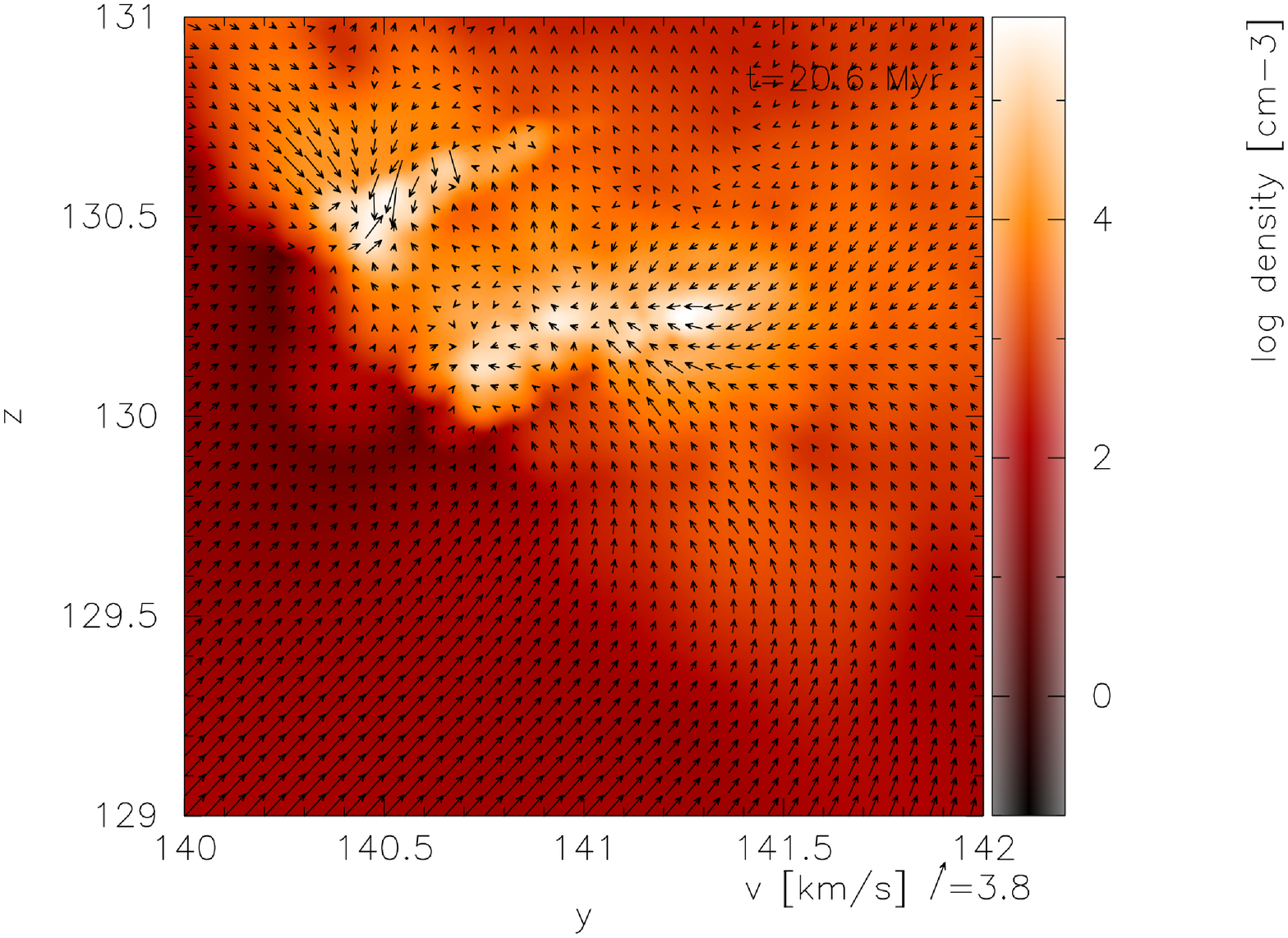}
\includegraphics[angle=0,scale=0.45]{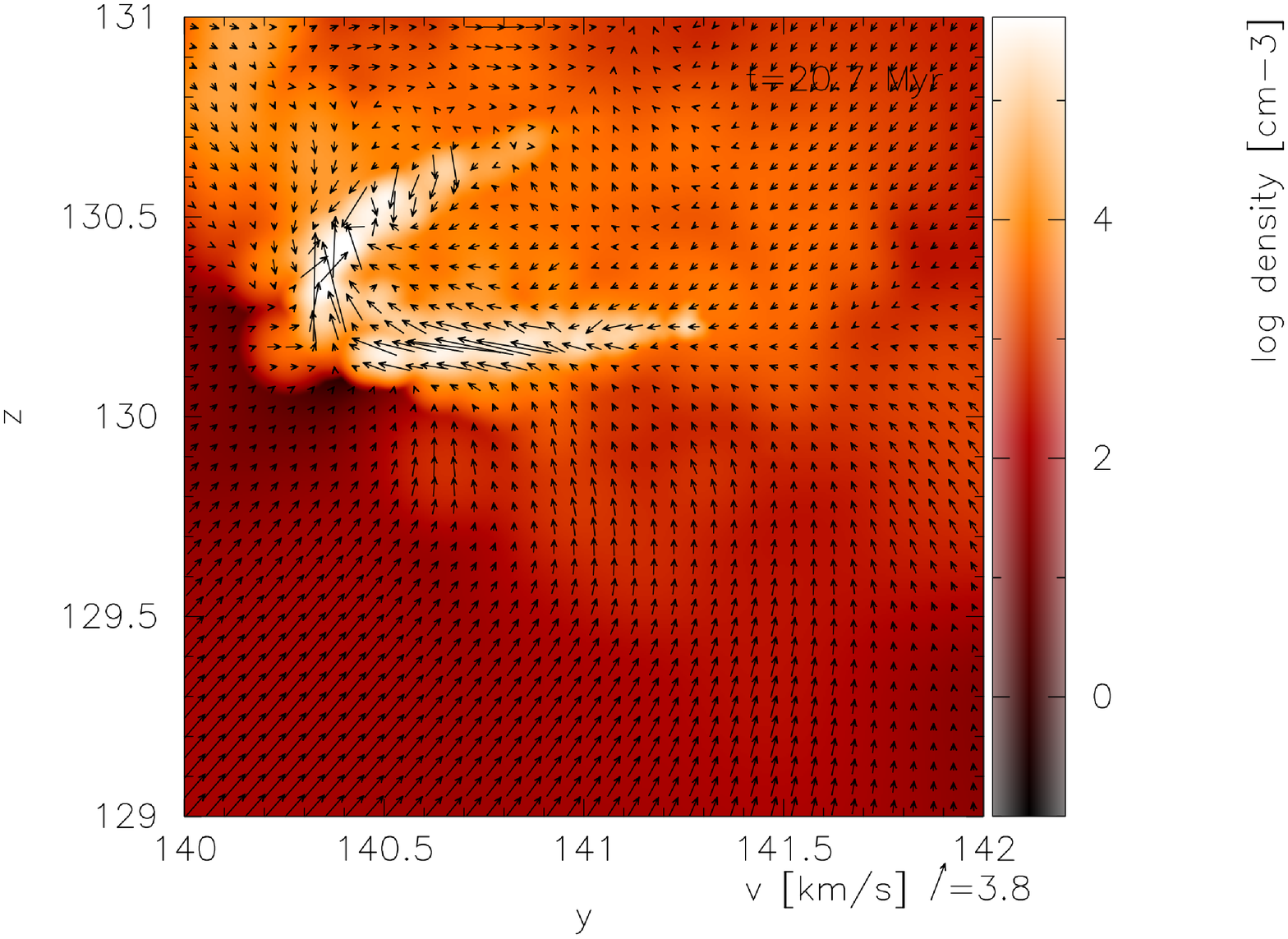}
\end{center}
\caption{
Snapshots of molecular filaments merging with each other extracted from the simulation presented 
by V\'azquez-Semadeni et al. (2009). The time interval between the first ({\it top}) and second 
({\it bottom}) frames is 0.133 Myr. The units of the $z$- and $y$-axes are pc. 
The color scale shows the volume density (cm$^{-3}$) in a slice passing through the merging filaments. 
The arrows indicate the velocity of the gas in the $y$--$z$ plane. In the second frame ({\it bottom}),  
the two merged filaments have velocities roughly opposing each other by a few $\kms$. 
}
\label{fig10}
\end{figure}


\begin{thebibliography}{}

\bibitem[Anglada et al.(1998)]{Angla98} 
Anglada, G., Villuendas, E., Estalella, R., Beltr\'an, M. T., Rodr{\'\i}guez, L. F., Torrelles, J. M., 
\& Curiel, S. 1998, AJ, 116, 2953

\bibitem[Araya et al.(2005)]{Araya05} 
Araya, E., Hofner, P., Kurtz, S., Bronfman, L., \& DeDeo, S. 
2005, ApJS, 157, 279 


\bibitem[Ballesteros-Paredes et al.(1999)]{BP99} 
Ballesteros-Paredes, J., Hartmann, L., \& V\'azquez-Semadeni, E. 1999, ApJ, 527, 285 
	

\bibitem[Beltr\'an et al.(2006)]{Bel06} Beltr\'an, M. T., Cesaroni, R., Codella, C., Testi, L., 
Furuya, R. S., \& Olmi, L. 2006, Nature, 443, 427

\bibitem[Beuther et al.(2007)]{Beu07} 
Beuther, H., Churchwell, E. B., McKee, C. F., \& Tan, J. C. 2007, in 
Protostars and Planets V, ed. B. Reipurth, D. Jewitt, \& K. Keil (Tucson, AZ: Univ. Arizona Press), 
165

\bibitem[Bonnell et al.(2003)]{Bonn03} 
Bonnell, I. A., Bate, M. R., \& Vine, S. G. 2003, MNRAS, 343, 413 


\bibitem[Brooks et al.(2007)]{Brooks07}	
Brooks, K. J., Garay, G., Voronkov, M., \& Rodr\'iguez, L. F. 
2007, ApJ, 669, 459 

 
\bibitem[Bunn et al.(1995)]{Bunn95} 
Bunn, J. C., Hoare, M. G., \& Drew, J. E. 1995, MNRAS, 272, 346

 
\bibitem[Carrasco-Gonz\'alez et al.(2010)]{Carrasco10}
Carrasco-Gonz\'alez, C., Rodr\'iguez, L. F., Torrelles, J. M., Anglada, G., \& Gonz\'alez-Mart\'in, O. 
2010, AJ, 139, 2433 
 


\bibitem[Cesaroni et al.(1999)]{Cesa99} 
Cesaroni, R., Felli, M., Jenness, T., Neri, R., Olmi, L., Robberto, M., Testi, L., \& Walmsley, C. M. 
1999, A\&A, 345, 949 

\bibitem[Cesaroni et al.(2005)]{Cesa05} 	
Cesaroni, R., Neri, R., Olmi, L., Testi, L., Walmsley, C. M., \& Hofner, P. 
2005, A\&A, 434, 1039

 
\bibitem[Cyganowski et al.(2008)]{Cyga08} 
Cyganowski, C. J. et al. 2008, AJ, 136, 2391


\bibitem[Davies et al.(2010)]{Davies10}	
Davies, B., Lumsden, S. L., Hoare, M. G., Oudmaijer, R. D., \& de Wit, W. 
2010, MNRAS, 402, 1504

\bibitem[de Wit et al.(2007)]{dW07} 
de Wit, W. J., Hoare, M. G., Oudmaijer, R. D., \& Mottram, J. C. 
2007, ApJ, 671, L169 

\bibitem[de Wit et al.(2010)]{dW09}
de Wit, W. J., Hoare, M. G., Oudmaijer, R. D., \& Lumsden, S. L. 
2010, A\&A, 515, 45


\bibitem[Franco-Hern\'andez et al.(2009)]{FH09}	
Franco-Hern\'andez, R., Moran, J. M., Rodr\'iguez, L. F., \& Garay, G. 2009, ApJ, 701, 974

\bibitem[Franco-Hern\'andez \& Rodr\'iguez(2004)]{FHR04} 
Franco-Hern\'andez, R., \& Rodr\'iguez, L. F. 2004, ApJ, 604, L105 


\bibitem[Franco et al.(2000)]{Franco00} Franco, J., Kurtz, S., Hofner, P., Testi, L., 
Garc{\'\i}a-Segura, G., \& Martos, M. 2000, ApJ, 542, L143

\bibitem[Galv\'an-Madrid et al.(2009)]{GM09} 
Galv\'an-Madrid, R., Keto, E., Zhang, Q., Kurtz, S., Rodr\'iguez, L. F., \& Ho, P. T. P. 
2009, ApJ, 706, 1036 

\bibitem[Galv\'an-Madrid et al.(2008)]{GM08} 
Galv\'an-Madrid, R., Rodr{\'\i}guez, L. F., Ho, P. T. P., \& Keto, E. 2008, ApJ, 674, L33

 
\bibitem[Gibb \& Hoare(2007)]{GH07} 
Gibb, A. G., \& Hoare, M. G. 2007, MNRAS, 380, 246 


\bibitem[Heitsch et al.(2008)]{Heit08} 
Heitsch, F., Hartmann, L. W., Slyz, A. D., Devriendt, J. E. G., \& Burkert, A. 
2008, ApJ, 674, 316 

\bibitem[Ho et al.(2004)]{Ho04} Ho, P. T. P., Moran, J. M., \& Lo, K. Y. 2004, ApJ, 616, L1

 
\bibitem[Hoare et al.(2007)]{Hoare07}
Hoare, M. G., Kurtz, S. E., Lizano, S., Keto, E., \& Hofner, P. 
2007, in Protostars and Planets V, ed. B. Reipurth, D. Jewitt, \& K. Keil 
(Tucson, AZ: Univ. Arizona Press), 181


 
\bibitem[Hofner et al.(2007)]{Hofner07} 
Hofner, P., Cesaroni, R., Olmi, L., Rodr\'iguez, L. F., Mart\'i, J., \& Araya, E. 
2007, A\&A, 465, 197 


\bibitem[Jaffe et al.(1982)]{Jaffe82} Jaffe, D. T., Stier, M. T., \& Fazio, G. G. 
1982, ApJ, 252, 601 

\bibitem[Jappsen et al.(2005)]{Jap05} Jappsen, A.-K.,
Klessen, R.~S., Larson, R.~B., Li, Y., \& Mac Low, M.-M.\ 2005, \aap,
435, 611 

 
\bibitem[Jim\'enez-Serra et al.(2010)]{JS10}
Jim\'enez-Serra, I., Caselli, P., Tan, J. C., Hernandez, A. K., Fontani, F., Butler, M. J., \& van Loo, S. 
2010, MNRAS, 406, 187 


\bibitem[Jim\'enez-Serra et al.(2009)]{JS09} 
Jim\'enez-Serra, I., Mart\'in-Pintado, J., Caselli, P., Mart\'in, S., Rodr\'iguez-Franco, A., 
Chandler, C., \& Winters, J. M. 2009, ApJ, 703, L157 

\bibitem[Keto(2003)]{Keto03} Keto, E. 2003, ApJ, 599, 1196


\bibitem[Keto \& Wood(2006)]{KW06} Keto, E., \& Wood, K. 2006, ApJ, 637, 850

\bibitem[Keto et al.(2008)]{KZK08} Keto, E., Zhang, Q., \& Kurtz, S. 2008, ApJ, 672, 423

\bibitem[Klaassen et al.(2009)]{Klaass09} 	
Klaassen, P. D., Wilson, C. D., Keto, E. R., \& Zhang, Q.  2009, ApJ, 703, 1308 

\bibitem[Kratter \& Matzner(2009)]{KraMa09} 
Kratter, K. M., \& Matzner, C. D. 2006, MNRAS, 373, 1563 

\bibitem[Krumholz et al.(2009)]{Krum09} 
Krumholz, M. R., Klein, R. I., McKee, C. F., Offner, S. S. R., Cunningham, A. J. 
2009, Science, 323, 754 

\bibitem[Kurtz et al.(2000)]{Kurtz00} 
Kurtz, S., Cesaroni, R., Churchwell, E., Hofner, P., \& Walmsley, C. M. 2000, 
Protostars and Planets IV, ed. V. Mannings, A. P. Boss, \& S. S. Russell 
(Tucson, AZ: Univ. Arizona Press), 299


 
\bibitem[Lovas(2004)]{Lovas04}
Lovas, F.J., 2004, J. Phys. Chem. Ref. Data, 33, 177 


 
\bibitem[M\"{u}ller et al.(2005)]{Muller05}
M\"{u}ller, H. S. P., Schl\"{o}der, F., Stutzki, J., \& Winnewisser, G. 
2005, J. Mol. Struct., 742, 215


\bibitem[Ossenkopf \& Henning(1994)]{OH94} 
Ossenkopf, V., \& Henning, T. 1994, A\&A, 291, 943 

\bibitem[Patel et al.(2005)]{Patel05} 
Patel, N., et al. 2005, Nature, 437, 109 

\bibitem[Peters et al.(2010)]{Peters10} 
Peters, T., Banerjee, R., Klessen, R. S., Mac Low, M.-M., Galv\'an-Madrid, R., \& Keto, E. 
2010, ApJ, 711, 101

\bibitem[Qiu et al.(2009)]{Qiu09} 
Qiu, K., Zhang, Q., Wu, J., \& Chen, H.-R. 2009, ApJ, 696, 66 


\bibitem[Remijan et al.(2004)]{Remij04} Remijan, A., Sutton, E. C., Snyder, L. E., Friedel, D. N., 
Liu, S.-Y., \&  Pei, C.-C. 2004, ApJ, 606, 917

\bibitem[Rengarajan \& Ho(1996)]{RH96} 
Rengarajan, T. N., \& Ho, P. T. P. 1996, ApJ, 465, 363 



\bibitem[Rod\'on et al.(2008)]{Rodon08} 
Rod\'on, J. A., Beuther, H., Megeath, S. T., \& van der Tak, F. F. S. 2008, A\&A, 490, 213 

\bibitem[Rodr\'iguez et al.(2005)]{Rod05} 
Rodr\'iguez, L. F., Garay, G., Brooks, K. J., \& Mardones, D.  2005, ApJ, 626, 953 

\bibitem[Rodr\'iguez et al.(2008)]{Rod08} 
Rodr\'iguez, L. F., Moran, J. M., Franco-Hern\'andez, R., Garay, G., Brooks, K. J., \& Mardones, D. 
2008, AJ, 135, 2370

\bibitem[Rosolowsky et al.(2008)]{Roso08} 
Rosolowsky, E. W., Pineda, J. E., Foster, J. B., Borkin, M. A., Kauffmann, J., Caselli, P., Myers, P. C., 
\& Goodman, A. A.  2008, ApJS, 175, 509 


\bibitem[Shepherd et al.(2001)]{Shep01} 
Shepherd, D. S., Claussen, M. J., \& Kurtz, S. 2001, Science, 292, 1513 

\bibitem[Smith et al.(2009)]{Smith09} 
Smith, R. J., Longmore, S., \& Bonnell, I. 2009, MNRAS, 400, 1775 

\bibitem[Springel et al.(2001)]{Spring01} Springel, V., Yoshida, 
N., \& White, S.~D.~M.\ 2001, New Astronomy, 6, 79 

\bibitem[Stier et al.(1984)]{Stier84} 
Stier, M. T., Jaffe, D. T., Rengarajan, T. N., Fazio, G. G., Maxson, C. W., McBreen, B., 
Loughran, L., Serio, S., \& Sciortino, S. 1984, ApJ, 283, 573

\bibitem[V\'azquez-Semadeni et al.(2007)]{VS07} 
V{\'a}zquez-Semadeni, E., G{\'o}mez, G.~C., Jappsen, A.~K., 
Ballesteros-Paredes, J., Gonz{\'a}lez, R.~F., 
\& Klessen, R.~S.\ 2007, ApJ, 657, 870 

\bibitem[V\'azquez-Semadeni et al.(2009)]{VS09} 
V\'azquez-Semadeni, E., G\'omez, G. C., Jappsen, A. K.,
Ballesteros-Paredes, J., \& Klessen, R. S.  2009, ApJ, 707, 1023

\bibitem[van der Tak \& Menten(2005)]{vdTM05} 
van der Tak, F. F. S., \& Menten, K. M. 2005, A\&A, 437, 947 

\bibitem[van der Tak et al.(2005)]{vdT05} 
van der Tak, F. F. S., Tuthill, P. G., \& Danchi, W. C. 2005, A\&A, 431, 993 

\bibitem[van der Tak et al.(2000)]{vdT00} 
van der Tak, F. F. S., van Dishoeck, E. F., Evans, N. J. II, \& Blake, G. A. 
2000, ApJ, 537, 283

\bibitem[Westerhout(1958)]{W58} Westerhout, G. 
1958, Bulletin of the Astronomical Institutes of the Netherlands, 14, 215 

\bibitem[Williams et al.(2009)]{Will09} 
Williams, J. P., Mann, R. K., Beaumont, C. N., Swift, J. J., Adams, J. D., Hora, J., Kassis, M., 
Lada, E. A., \& Rom\'an-Z\'u\~niga, C. G. 	2009, ApJ, 699, 1300 

\bibitem[Wilner et al.(1994)]{Wil94} Wilner, D. J., Wright, M. C. H., \& Plambeck, R. L. 1994, ApJ, 
422, 642

\bibitem[Wu \& Evans(2003)]{Wu03} Wu, J., \& Evans, N. J., II 
2003, ApJ, 592, L79

\bibitem[Zapata et al.(2009)]{Zap09} 
Zapata, L. A., Ho, P. T. P., Schilke, P., Rodr\'iguez, L. F., Menten, K., Palau, A., Garrod, R. T. 
2009, ApJ, 698, 1422 

\bibitem[Zapata et al.(2008)]{Zap08} 
Zapata, L. A., Palau, A., Ho, P. T. P., Schilke, P., Garrod, R. T., Rodr\'iguez, L. F., \&  Menten, K. 
2008, A\&A, 479 L25 


\bibitem[Zhang et al.(2007)]{Zhang07} 
Zhang, Q., Hunter, T. R., Beuther, H., Sridharan, T. K., Liu, S.-Y., Su, Y.-N., Chen, H.-R., \& Chen, Y. 
2007, ApJ, 658, 1152 

\bibitem[Zhang et al.(1998)]{Zhang98b}
Zhang, Q., Hunter, T. R., \& Sridharan, T. K. 	
1998, ApJ, 505, L151 


\bibitem[Zhang et al.(2009)]{Zhang09} 
Zhang, Q., Wang, Y., Pillai, T., \& Rathborne, J. 2009, ApJ, 696, 268 

\bibitem[Zinnecker \& Yorke(2007)]{ZY07} 
Zinnecker, H., \& Yorke, H. W. 2007, ARA\&A, 45, 481 

\end{thebibliography}
\end{document}